\documentclass[prb,aps,twocolumn,superscriptaddress,showpacs]{revtex4}

\usepackage{graphicx}

\begin{document}

\title{From incommensurate to dispersive spin-fluctuations: The high-energy inelastic spectrum in superconducting YBa$_{2}$Cu$_{3}$O$_{6.5}$}

\author{C. Stock}
\affiliation{Department of Physics, University of Toronto, Ontario, Canada M5S 1A7}
\author{W. J. L. Buyers}
\affiliation{National Research Council, Chalk River, Ontario, Canada K0J 1JO}
\affiliation{Canadian Institute of Advanced Research, Toronto, Ontario, Canada M5G 1Z8}
\author{R. A. Cowley}
\affiliation{Oxford Physics, Clarendon Laboratory, Oxford, United Kingdom OX1 3PU}
\author{P. S. Clegg}
\affiliation{Department of Physics, University of Toronto, Ontario, Canada
M5S 1A7}
\author{R. Coldea}
\affiliation{Oxford Physics, Clarendon Laboratory, Oxford, United Kingdom OX1 3PU}
\author{C. D. Frost}
\affiliation{ISIS Facility, Rutherford Appleton Laboratory, Didcot, Oxon, United Kingdom OX11 0QX}
\author{R. Liang}
\affiliation{Physics Department, University of British Columbia, Vancouver, B. C., Canada V6T 2E7}
\affiliation{Canadian Institute of Advanced Research, Toronto, Ontario, Canada M5G 1Z8}
\author{D. Peets}
\affiliation{Physics Department, University of British Columbia, Vancouver, B. C., Canada V6T 2E7}
\author{D. Bonn}
\affiliation{Physics Department, University of British Columbia, Vancouver, B. C., Canada V6T 2E7}
\affiliation{Canadian Institute of Advanced Research, Toronto, Ontario, Canada M5G 1Z8}
\author{W. N. Hardy}
\affiliation{Physics Department, University of British Columbia, Vancouver, B. C., Canada V6T 2E7}
\affiliation{Canadian Institute of Advanced Research, Toronto, Ontario, Canada M5G 1Z8}
\author{R. J. Birgeneau}
\affiliation{Department of Physics, University of Toronto, Ontario, Canada M5S 1A7}
\affiliation{Canadian Institute of Advanced Research, Toronto, Ontario, Canada M5G 1Z8}

\date{\today}

\begin{abstract}

    We have investigated the spin fluctuations at energy transfers up to $\sim$ 110 meV, well above the resonance energy (33 meV) in the YBa$_{2}$Cu$_{3}$O$_{6.5}$ ortho-II superconductor using neutron time-of-flight and triple-axis techniques.  The spectrum at high energies differs from the low-energy incommensurate modulations previously reported where the incommensurate wave vector is largely independent of energy.  Well above the resonance the peak of the spin response lies at wave vectors that increase with energy.  Within error the excitations at all energies above the resonance are best described by a ring around the ($\pi$, $\pi$) position. The isotropic wave-vector pattern differs from a recently reported square pattern in different but related systems.  The spin spectral weight at high-energies is similar to that in the insulator but the characteristic velocity is $\sim$ 40\% lower.  We introduce a method of extracting the acoustic and optic weights at all energies from time-of-flight data.  We find that the optic spectral weight extends to surprisingly low-energies of $\sim$ 25 meV, and infer that the bilayer spin correlations weaken with increase in hole doping.  When the low-energy optic excitations are taken into account we measure the total integrated weight around ($\pi$, $\pi$), for energies below 120 meV, to agree with that expected from the insulator.  As a qualitative guide, we compare spin-wave calculations for an ordered and a disordered stripe model and describe the inadequacy of this and other stripe models for the high-energy fluctuations.

\end{abstract}


\maketitle

\section{Introduction}

    The discovery of superconductivity in the cuprates has resulted in a vast amount of interest in the spin fluctuations of these nearly two-dimensional materials.~\cite{Manousakis91:63,Kastner98:70}  Despite the intense research the cuprates have received, there is no generally accepted theory of the mechanism for the high temperature superconductivity.  Due to the interplay between antiferromagnetism and superconductivity the spin fluctuations are drastically altered upon doping the parent insulator and are also strongly correlated with superconductivity.  Understanding the spin-fluctuations in the cuprates is an important step to understanding superconductivity.

    We have recently completed an extensive experimental study of the temperature, momentum, and energy dependence of the spin fluctuations below 40 meV, including the resonance peak at $\sim$ 33 meV, for the partially detwinned YBa$_{2}$Cu$_{3}$O$_{6.5}$ ortho-II superconductor.~\cite{Stock04:69}  The good structural properties from the oxygen chain order removes the effect of impurity scattering of the quasiparticles and spin fluctuations due to disorder.  We found the low-energy excitations to be gapless and to give rise to incommensurate peaks only displaced along the [100] direction from the ($\pi$, $\pi$) position.  This contrasts with the symmetric cone of scattering which disperses from the ($\pi$, $\pi$) position observed in the parent insulating materials and illustrates the need for studies to be conducted on detwinned samples so that the $a-b$ anisotropy can be unambiguously resolved. At $\sim$ 33 meV we found a commensurate resonance peak which was strongly enhanced in the superconducting state.  One of the surprising aspects of the resonance peak is that, despite the low energy scale of superconductivity (T$_{c}$$\sim$59 K$\sim$ 5 meV), superconductivity has a very strong effect on the high-energy excitations around the resonance.   An important question is then how the excitations above the resonance energy are affected by superconductivity and hole doping.

    The nature of the high-energy spin fluctuations has previously been studied only in the parent insulating cuprate compounds.  Using spallation neutrons Hayden \textit{et al.}~\cite{Hayden96:54} were able to map both the acoustic and optic modes in YBCO$_{6.15}$ and found that the results were well described by linear spin-wave theory.  They found a large $\sim$ 75 meV energy gap for the optic fluctuations, an in-plane exchange constant of $\sim$ 125 meV, and a bilayer coupling of $\sim$ 11 meV. The dispersion was found to extend up to $\sim$ 250 meV.  These results have also been confirmed by reactor measurements by Reznik \textit{et al.}~\cite{Reznik96:53}  The entire spin-wave dispersion of the single layer compound, La$_{2}$CuO$_{4}$, has been studied in detail by Coldea \textit{et al.}~\cite{Coldea01:86}

    After this paper was submitted, there have been new developments because of letters to Nature by Hayden \textit{et al.}~\cite{Hayden04:429} and Tranquada \textit{et al.}~\cite{Tranquada04:429} on the high-energy spin fluctuations in YBCO$_{6.6}$ and LBCO.  We have therefore included discussion and comparison with the square-like pattern in these cuprates and will show that the spin fluctuations on ordered YBCO$_{6.5}$ are quite different and specifically, they are isotropic in wave vector about $\bf{Q}$=($\pi$, $\pi$).
    
    The high-energy excitations of a doped superconducting sample were first measured by Bourges \textit{et al.}~\cite{Bourges97:56} in a twinned crystal of YBCO$_{6.5}$.  They found a broadening in $\mid$$\bf{Q}$$\mid$ of the correlated peak (at $\bf{Q}$=($\pi$, $\pi$)) with increasing energy.  Based on this result it was suggested that the spin fluctuations disperse (at least at high-energies) in a manner similar to that of the parent insulating compound.   The momentum dispersion of the magnetic excitations around the resonance was later studied in detail by Arai \textit{et al.}~\cite{Arai99:83} on, again, a twinned YBCO$_{6.7}$ sample.  They found evidence for two modes which meet at the resonance energy, one of which opens downwards and gives the incommensurate scattering measured by other groups,~\cite{Mook00:404} and a new mode which opens upwards in energy.  A similar result has also been suggested by a recent study on optimally doped YBCO$_{6.95}$ by Reznik \textit{et al.}~\cite{Reznik03:7591}  Another, recent experiment, on twinned La$_{1.875}$Ba$_{0.125}$CuO$_{4}$ has found high-energy excitations which are consistent with the dispersion of two-leg ladders (when S($\bf{Q}$, $\omega$) is averaged over domains).~\cite{Tranquada04:429}  Because all of the samples studied have been twinned, it is not clear if the high-energy scattering results from incommensurate peaks displaced along a particular direction (as we have observed in our reactor low-energy study) or if the excitations form a symmetric cone which is similar to that of the parent insulator (see the lower panel of Fig. \ref{acoustSFandcutintegration} for an illustration).  It is clearly important to compare the lineshape of the magnetic scattering above and below the resonance energy on a detwinned sample to resolve how the one-dimensional incommensurate scattering evolves to high-energies.  Our present study on a largely detwinned sample enables the intrinsic nature of the magnetic scattering at high-energies to be determined free of scattering by structural disorder.

    The nature of the spin susceptibility above the resonance energy has many important implications for the theory of the cuprates.  Recent effective band calculations using the random-phase approximation predict the presence of spin excitations that have a velocity much higher than that of the parent insulator.~\cite{Kao00:61} In contrast, linear spin-wave models based on a stripe ground state predict that the excitations at high energies are nearly symmetric around the ($\pi$, $\pi$) position and disperse almost linearly with energy.~\cite{Carlson04:2231,Kruger03:67,Kruger04:1354,Batista01:64}  Quantum critical theories of two-dimensional Heisenberg antiferromagnets suggest that the resonance energy can be interpreted as a gap which softens as a critical point is approached.  Such a theory would suggest that the excitations well above the resonance are qualitatively similar to those of the parent insulating compound.~\cite{Chubukov94:49}  Therefore, a detailed knowledge of the true spin spectrum (involving dispersion and lineshape) above the resonance energy is crucial in understanding the cuprates and in constraining theories.

    We present a high-energy neutron inelastic study on a well-ordered superconducting YBCO sample.  We investigate the momentum and energy dependence of the magnetic excitations above the resonance energy.  This study complements our previous work on the low-energy spectrum using a reactor neutron source.~\cite{Stock04:69} The paper is divided into two sections and a discussion.  The first section describes the experimental setup.  The second section describes how we extracted the acoustic and optic weights using the known bilayer structure factors as well as the dispersion at high-energies.  We discuss the integrated intensity in terms of the total moment sum rule.  In the Appendix we describe linear spin-wave calculations for an ordered and disordered stripe ground state and compare the results qualitatively to the experimental data.

\section{Experimental Details}

\begin{figure}[t]
\includegraphics[width=8cm] {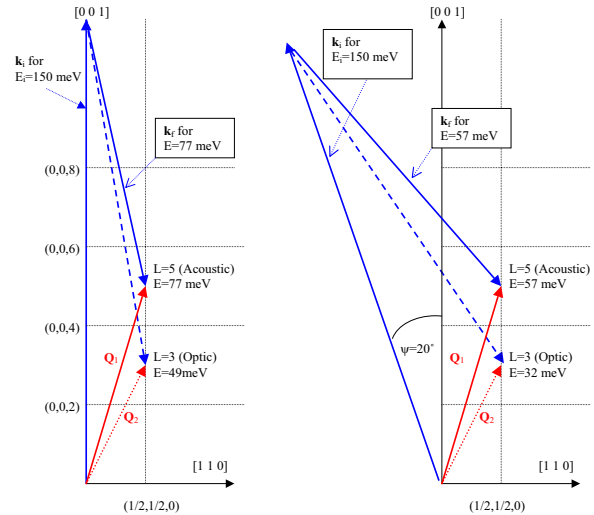}
\caption{\label{kinematics}  (Color online) The scattering diagram for a constant in-plane wave vector of (H,H)=(1/2,1/2) shows that as the energy transfer, E, varies, so does the c-axis component of wave vector L(E), thus enabling the acoustic (odd) and optic (even) symmetries of the spin response excitations to be determined by combining data.   In the left panel the 150 meV incident neutron wave vector $\bf{k_{i}}$ is brought in along the [0 0 1] direction. The primarily acoustic component of the response at E=77 meV (solid lines) is obtained near L=5 as well as the optic response at 49 meV near L=3. To obtain a different relation between E and L, so as to determine the ratio of acoustic to weight at different energies, the incident beam is brought in at variety of angles, $\psi$, to the [0 0 1] crystal axis (right panel). }  
\end{figure}

    The sample consisted of six orthorhombic crystals with total volume $\sim$ 6 cm$^{3}$ aligned on a multi-crystal mount with a combined rocking curve width of $\sim$ 1.5$^{\circ}$.  Details of the crystal growth, the detwinning by stress along the \textit{a} axis, and the oxygen order have been published earlier.~\cite{Stock04:69,Peets02:15}  The sample has previously been measured to be partially detwinned with the majority domain occupying 70 $\%$ of the total volume and with good oxygen order.  Oxygen correlation lengths exceed $\sim$ 100 \AA\ in the \textit{a} and \textit{b} directions.~\cite{corr} The DC magnetization shows a superconducting onset temperature of 59 K with a width of $\sim$ 2.5 K.  The low temperature lattice constants were measured to be \textit{a}=3.81 \AA, \textit{b}=3.86 \AA, and \textit{c}=11.67 \AA.

\begin{figure}[t]
\includegraphics[width=8cm] {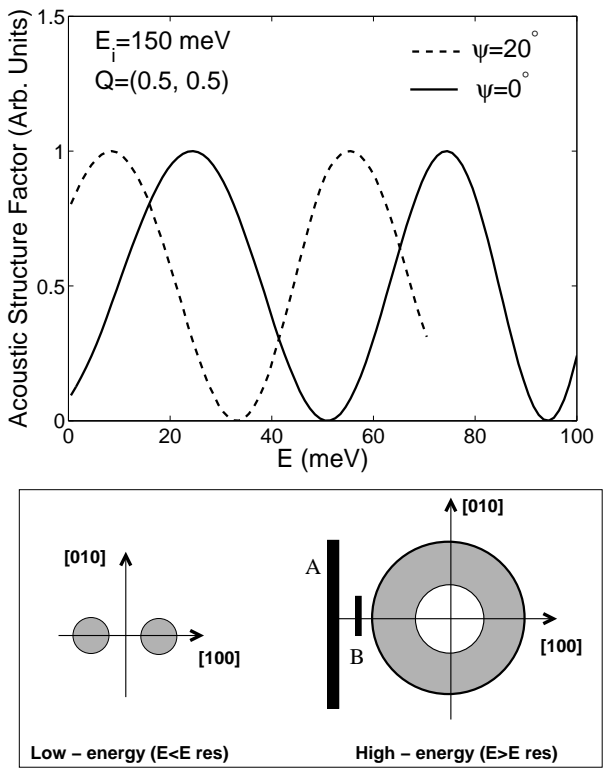}
\caption{\label{acoustSFandcutintegration} The upper panel shows the calculated acoustic structure factor at the $\bf{Q}$=(1/2, 1/2) position as a function of energy transfer.  It shows how one can choose a different fraction of acoustic and optic weight at at given energy transfer by changing the angle ($\psi$) between the incident beam direction and the c-axis.  The lower panel shows, schematically, the observed difference between scattering above and below the resonance energy.  The scattering below the resonance energy consists of two peaks displaced along the [100] direction while the high-energy scattering comprises a ring.  The arrows illustrate how cuts were done to study the dispersion of the correlated peaks.  The filled boxes (denoted A and B) represent the integration windows used to study the spectral weight and the lineshape as a function of energy transfer as discussed in the text.}
\end{figure}

    Neutron scattering experiments were conducted by means of time-of-flight and triple-axis techniques.  Time-of-flight measurements were made at the MAPS spectrometer at the ISIS spallation source. The MAPS position-sensitive-detectors cover 16 m$^{2}$ area divided into 36864 approximately square pixels.  The position-sensitive-detectors are designed so that there is no direction that has significantly poorer $\bf{Q}$ resolution arising from the shape of the detectors.  A monochromatic beam was produced by the use of a Fermi chopper spinning at a frequency of 400 Hz with the incident energy fixed at E$_{i}$=150 meV.  The sample was mounted with the [001] and [110] directions in the horizontal plane and with the incident beam ($\bf{k}_{i}$) initially aligned along the [001] direction.  Because the value of L, and hence the bilayer structure factor, varies with energy transfer (see Fig.\ref{kinematics} and the next section), this orientation gives only a limited ability to measure the acoustic and optic response independently.  In order to extract the acoustic and optic components for a range of energies (see Fig. \ref{acoustSFandcutintegration}), we varied the angle between the incident beam and the [001] axis, defined as $\psi$, with the rotation axis along [1$\overline{1}$0].  This procedure was chosen over varying the incident energy (used elsewhere~\cite{Arai99:83}) to avoid changes in the resolution function.  The full-width at half-maximum (FWHM) energy resolution was measured from the incoherent elastic scattering to be $\sim$ 8 meV.  Using this method we have been able to study the spectral weight and the dispersion of both the acoustic and optic modes up to high-energies.

    To calibrate the spectrometer, a standard vanadium sample with a known mass was used.  To obtain $\chi''(\bf{q}, \omega)$ we have taken the cross section to be isotropic in spin with  definitions of $\chi''$ and the total cross-section that we used previously~\cite{Stock04:69} and that agree with those of other groups.~\cite{Hayden98:241}  The definitions used here are identical to that used in Ref. \onlinecite{Hayden96:54} and in Ref. \onlinecite{Fong00:61}.  The definition of $\chi$$'$$'$ used here differs to that used in Ref. \onlinecite{Bourges97:79} such that our values of $\chi$$'$$'$ are 1.5 times greater than that in  Ref. \onlinecite{Bourges97:79}.  As a consistency check (and as discussed later) we have found that the total integral ($\int d^{3}q \int d\omega$) of the low-energy scattering obtained from the vanadium calibration agrees well with that previously obtained~\cite{Stock04:69} using an internal phonon calibration.

    Triple-axis measurements were conducted at the DUALSPEC spectrometer located at the C5 beam of the NRU reactor at Chalk River Laboratories.  A focusing graphite (002) monochromator and a graphite (002) analyzer were used.  A pyrolytic graphite filter in the scattered beam eliminated higher order reflections and the final energy was fixed at E$_{f}$=14.6 meV.  The horizontal collimation was set at [33$'$ 29$'$ \textit{S} 51$'$ 120$'$].  The vertical collimation was fixed at [80$'$ 240$'$ \textit{S} 214$'$ 430$'$].  The six-crystal assembly was mounted in a closed-cycle refrigerator and aligned such that reflections of the form $(HHL)$ lay in the scattering plane.  To put these measurements on an absolute scale, a transverse acoustic phonon was measured around the (0 0 6) position.  This was the same procedure used previously.~\cite{Stock04:69}

\section{Results}

    In magnetic systems that are two-dimensional (eg. La$_{2}$CuO$_{4}$), or one-dimensional (eg. CsNiCl$_{3}$), the non-dispersive nature of the excitations along one or two directions makes the time-of-flight technique well suited for mapping out the dispersion of the excitations.  For such cases the non-dispersive momentum direction is typically placed parallel to the incident wave-vector such that the scattering can be projected on the plane perpendicular to $\bf{k}_{i}$ and the entire dispersion (in q$_{x}$ and q$_{y}$) can be mapped out.

    This technique has been successfully applied to the single-layer La$_{2}$CuO$_{4}$ insulator where the excitations are nearly two-dimensional and there is little dispersion along the [001] axis.  The situation in YBa$_{2}$Cu$_{3}$O$_{6+x}$ is complicated by the presence of two strongly coupled CuO$_{2}$ layers in the unit cell.  There is a weak antiferromagnetic coupling (of order 10 meV) between the two layers (separation $d$) in each unit cell. This means that the spin excitations can be categorized as odd (or acoustic) where the spin fluctuations are out of phase between the layers, or even (or optic) where they are in phase.~\cite{Tranquada89:40}  Because the bilayer coupling is antiferromagnetic the optic spin waves in the insulator have a larger energy than the acoustic.  

    The acoustic mode intensity varies along the [001] as sin$^{2}(Q_{z}d/2)$, where $d$ is the distance between the two layers in the unit cell, while the optic mode strength is proportional to cos$^{2}(Q_{z}d/2)$.~\cite{Tranquada89:40}  Therefore, for a fixed experimental configuration different L values, and hence different acoustic and optic weights, will be sampled because L varies with energy transfer.  The reason for L varying with energy is schematically outlined in Fig. \ref{kinematics} which shows the scattering triangle for two different energy transfers, and hence two different final energies (note that E$_{i}$ is fixed in this experiment).  For a particular energy transfer corresponding to a final energy E$_{f}$, the constraint of having the momentum transfer of the form $\bf{Q}$=(1/2, 1/2, L) fixes the value of L.  Therefore, by changing the energy transfer, but keeping $\bf{Q}$ to have the form $\bf{Q}$=(1/2, 1/2, L), will result in L varying with energy transfer.

    The effect of this is further illustrated in the upper panel of Fig. \ref{acoustSFandcutintegration}, which shows the acoustic structure factor at $\bf{Q}$=(1/2, 1/2) as a function of energy transfer for $\psi$=0$^{\circ}$ and 20$^{\circ}$.  The acoustic structure factor varies along [001] as sin$^{2}(Q_{z}d/2)$, but since the value of Q$_{z}$ depends on the energy transfer for a fixed $(Q_{x}, Q_{y})$=(1/2, 1/2), different acoustic and optic weights will be sampled as a function of energy.  Therefore, for $\psi$=0 (see Fig. \ref{kinematics}), to sample purely acoustic scattering an energy transfer of about 30 meV or 75 meV is required.  Fig. \ref{acoustSFandcutintegration} also illustrates that varying $\psi$ moves the peak of the acoustic structure factor to different energy transfers.  In other studies, this problem has been overcome by tuning the incident energy so that the energy position of the maximum in the structure factors would shift to the required position.  Such a procedure inevitably alters the resolution function and therefore the relative weight of the acoustic and optic modes. As a result, the dispersion of the acoustic and optic modes cannot be reliably determined without the aid of resolution corrections.

    To avoid this, we kept the incident energy fixed and varied the angle $\psi$ between the [001] axis and {\bf{{k}$_{i}$}} (illustrated in Fig. \ref{kinematics}).  We have collected data for $\psi$=0, 20$^{\circ}$, and 30$^{\circ}$.  In each configuration we were able to extract data in up to 12 different Brillouin zones (including the ($\pm$1/2,$\pm$1/2), ($\pm$3/2,$\pm$1/2) and permutations) each of which occur for different values of L. The intensity of the magnetic scattering for each zone, sample orientation ($\psi$), and energy bin was then obtained by taking a cut (along the [1, 0] and [0, 1] directions) through the scattering at a constant energy transfer. This process relied on the MSLICE program.~\cite{mslice}

    In this section we discuss the experimental results in terms of four different subsections.  In the first section we discuss the lineshape of the high-energy acoustic scattering and its symmetry in the (H, K) plane.  This is based on two-dimensional slices integrating $\pm$7.5 meV in energy. The next section we study the momentum dependence of the scattering and its dispersion in detail.  This involves an analysis where we integrated over fine slices in momentum, (see Box \textit{B} in the lower panel of Fig. \ref{acoustSFandcutintegration}) and therefore do not smear out any lineshape changes as a function of energy transfer.  In the third section we describe the relative optic and acoustic weights as a function of energy transfer using an analysis in which we integrate over one momentum direction.  This is schematically represented by Box \textit{A} in the lower panel of Fig. \ref{acoustSFandcutintegration}.  This removes any detailed lineshape information and allows all energies to be treated equally.    In the final section we discuss the total integrated spectral weight and compare the results to other systems as well as the insulating compound (in particular YBCO$_{6.15}$).

\subsection{Spin Excitation Ring at High-energies}

\begin{figure}[t]
\includegraphics[width=8cm] {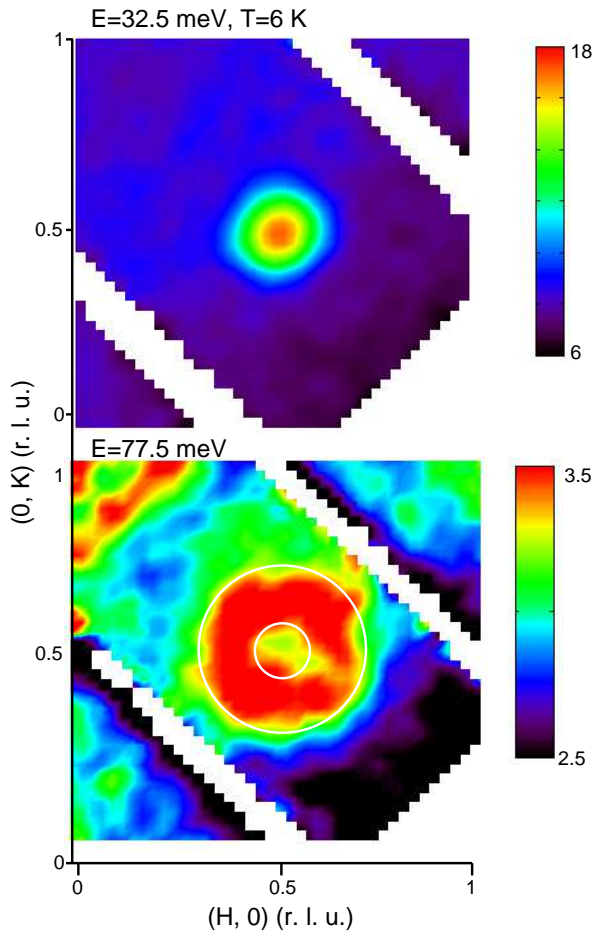}
\caption{\label{colour} (Color online) Smoothed two-dimensional slices through the correlated response for $\psi$=0 at 32.5 meV and 77.5 meV energy transfers and integrated $\pm$ 7.5 meV along the energy axis.  At 77.5 meV the ring of scattering shows the intersection with a constant energy surface of the cone of spin wave dispersion  emanating from the (1/2, 1/2) position.  Within statistics the velocity is isotropic.  The intensity represented by false color is in arbitrary units. Unsmoothed cuts of the data for various energy transfers are presented in Fig. \ref{symmetry} and Fig. \ref{acoustic_disp}.}
\end{figure}

\begin{figure}[t]
\includegraphics[width=8cm] {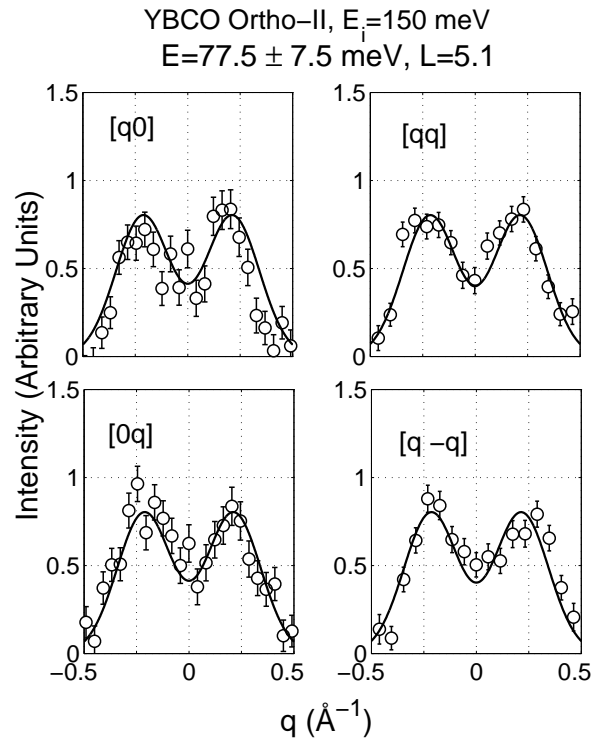}
\caption{\label{symmetry} One-dimensional cuts through the correlated ring at 77 meV energy transfer (displayed in Fig. \ref{colour}) along the [10], [01], [11] and [1$\overline{1}$] directions. The data was integrated $\pm$ 0.05 r.l.u. perpendicular to the cut direction and a sloping background was subtracted from each scan for comparison purposes.  The solid lines are Gaussians displaced equally from the ($\pi$, $\pi$) position.}
\end{figure}

    Examples of two-dimensional constant energy maps taken at $\psi$=0 are shown in Fig. $\ref{colour}$.  The two-dimensional slices shown in Fig. \ref{colour} have been averaged over the four ($\pm$ 1/2, $\pm$ 1/2) zones sampled by the MAPS detector array.  At the resonance energy of 33 meV the response is localized at ($\pi$, $\pi$).  Above the resonance at 78 meV a ring is observed showing the intersection of the constant energy surface with a two-dimensional cone of spin waves centered on ($\pi$, $\pi$) (this point is discussed more thoroughly in the next section).  The intensity around the ring is symmetric in the (H,K) plane as demonstrated in Fig. \ref{symmetry}.  Fig. \ref{symmetry} plots one-dimensional cuts through the ($\pi$, $\pi$) position at E=77 meV along the [1, 0], [0, 1], [1, 1], and [1, $\overline{1}$] directions.  For comparison purposes a sloping background has been subtracted from each cut.  The solid lines are Gaussians symmetrically displaced from the ($\pi$, $\pi$) position.  A comparison between the solid lines and the data shows that the scattering is consistent with an isotropic ring. Both the radius on wave vector and the intensity are independent of the direction in the a-b plane.

    Maps at all energies higher than the resonance are consistent with almost isotropic spin waves centered at ($\pi$, $\pi$).  Although an individual constant-energy contour, such as that at 77 meV in Fig. \ref{colour}, may show more counts at particular directions around the ring, our results are statistically consistent with a spin-wave dispersion whose velocity and spectral weight are isotropic in the (H, K) plane.  The line shape illustrated in the color map in Fig. \ref{colour} may also be compared with the diamond shape suggested by various calculations involving ladders.~\cite{Uhrig04:2659,Vojta04:2377} For the YBCO$_{6.5}$ system the scattered intensity clearly does not correspond to a square pattern nor to peaks displaced along [110] (see Fig. \ref{symmetry}).  There is no evidence at energies greater than 33 meV for the one-dimensional incommensurate modulation that exists below the resonance energy.~\cite{Stock04:69}

    Our observation that the scattering is symmetric is important in the context of a recent model proposed by Tranquada \textit{et al.} which suggests that the high-energy scattering in La$_{1.875}$Ba$_{0.125}$CuO$_{4}$ (LBCO) is similar to the excitations from ladders.~\cite{Tranquada04:429}  At high energies comparable to ours, the actual data in the color plots of Ref. \onlinecite{Tranquada04:429} seem to the eye to show less of the strong diamond-shaped anisotropy of the ladder model and are not dissimilar to the symmetric excitations we observe. We are able to resolve the excitation in wave vector because of the lower velocity in YBCO$_{6.5}$. 

    Recent calculations by Vojta and Ulbricht have found that a bond-centered stripe phase has an excitation spectrum very similar to that observed in LBCO.~\cite{Vojta04:2377}  For a ladder ground (or bond-centered stripe) state to give a symmetric excitation spectrum, equal contributions from horizontal and vertical ladders (or stripes) must be considered, as is present in a twinned superconductor.  Our results on partially detwinned YBCO$_{6.5}$ show that a detwinned structure does not give rise to a large anisotropy such as that expected for ladders.  This anisotropy should be particularly evident when comparing cuts along [11] to those along [10] or [01].  Ref. \onlinecite{Tranquada04:429} modelled the ladders for a twinned system (ladders along \textit{a} and \textit{b}) and so may give stronger scattering along [11] due to the overlap of the \textit{a} and \textit{b} domains.  It may be that the stripes (or ladders) are purely dynamic and therefore higher energy fluctuations do not sense the small orthorhombic nature of the crystal structure, even though the low energy fluctuations do.  We note that in the energy range studied in our experiment the dispersion relation for a ladder (like that observed in LBCO) is similar to the linear spin-waves observed in the insulator.  

    We will show later that the high-energy excitations have a similar dispersion and spectral weight to those of the parent insulator YBCO$_{6.15}$.  Therefore, incommensurate fluctuations that are displaced and are stronger along a particular direction occur only at low-energies indicating that the stripe texture barely influences the high-energy spin excitations.   This suggests that the high-energy excitations exhibit similar isotropy to those of an undoped insulator.   

    Our results differ from those reported in a recent study on twinned YBCO$_{6.6}$ by Hayden \textit{et al.}~\cite{Hayden04:429} who measured, not a ring of scattering, but four peaks displaced along the [110] directions.  For the YBCO$_{6.5}$ ortho-II detwinned system the scattered intensity clearly does not correspond to a square pattern nor to peaks along [110] (see Fig. \ref{symmetry}).  The different behavior may arise from the larger x=6.6 doping, for the low-energy incommensurate wave vector ($\delta$) in Ref. \onlinecite{Hayden04:429} is at least twice as large as that in Ortho-II YBCO$_{6.5}$.  Since $\delta$ scales approximately with doping, the YBCO$_{6.6}$ system of Ref. \onlinecite{Hayden04:429} must have a significantly larger effective doping in the CuO$_{2}$ plane than our Ortho-II YBCO$_{6.5}$.  We note that modestly larger T$_{c}$ of 63 K~\cite{Hayden04:429} is not a direct measure of doping since it is, in general, lowered by increased structural disorder.

\subsection{Dispersion of the Optic and Acoustic Modes}

    Despite the fact that the low-energy excitations below the resonance energy show one-dimensional incommensurate peaks slightly displaced along the [100] direction, the high energy excitations are close to being \textit{symmetric} around the ($\pi$, $\pi$) position.  This is illustrated in Fig. \ref{colour} which shows two-dimensional constant energy slices through the correlated peaks at T = 6 K. In particular, there is no preferred weight of the spin response at incommensurate wave vectors along $a^{*}$ relative to $b^{*}$.  Any modulation around the spin-wave cone lies within statistical error as demonstrated by the one-dimensional cuts displayed in Fig. \ref{symmetry}. Thus, the high-energy scattering forms a symmetric cone, whose momentum increases with increasing energy as expected for the linear, constant-velocity part of the spin-wave dispersion.  We have further observed that there is no qualitative change in the momentum dependence of the high-energy scattering in the normal state at T=85 K.  

\begin{figure}[t]
\includegraphics[width=8cm] {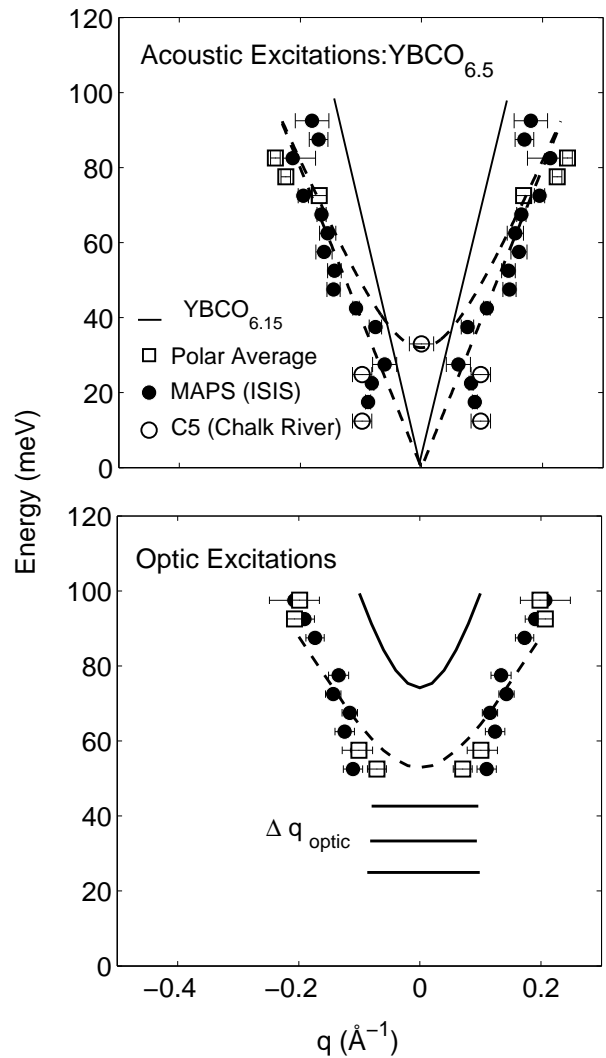}
\caption{\label{dispersion} The dispersion of the acoustic and optic modes with respect to the ($\pi$, $\pi$) position at T= 6 K is plotted.  The filled circles represent the peak positions of a two Gaussian fit to cuts along the [100] and [010] directions above the resonance energy and from cuts along only the [100] direction below the resonance energy.  The open circles represent the positions of the incommensurate peaks found in experiments conducted at Chalk River.  The open squares represent the peak position that result from a polar average around the ($\pi$,$\pi$) position.  The solid lines schematically represent the dispersion of the insulating compound as measured by Hayden \textit{et al.}  The dashed lines are fits of the high-energy dispersion to linear spin-wave theory for $\hbar \omega$ $>$ 40 meV as described in the text.  The horizontal bars show the $q$ width observed for three optic scans at constant energy.}
\end{figure}

\begin{figure}[t]
\includegraphics[width=8cm] {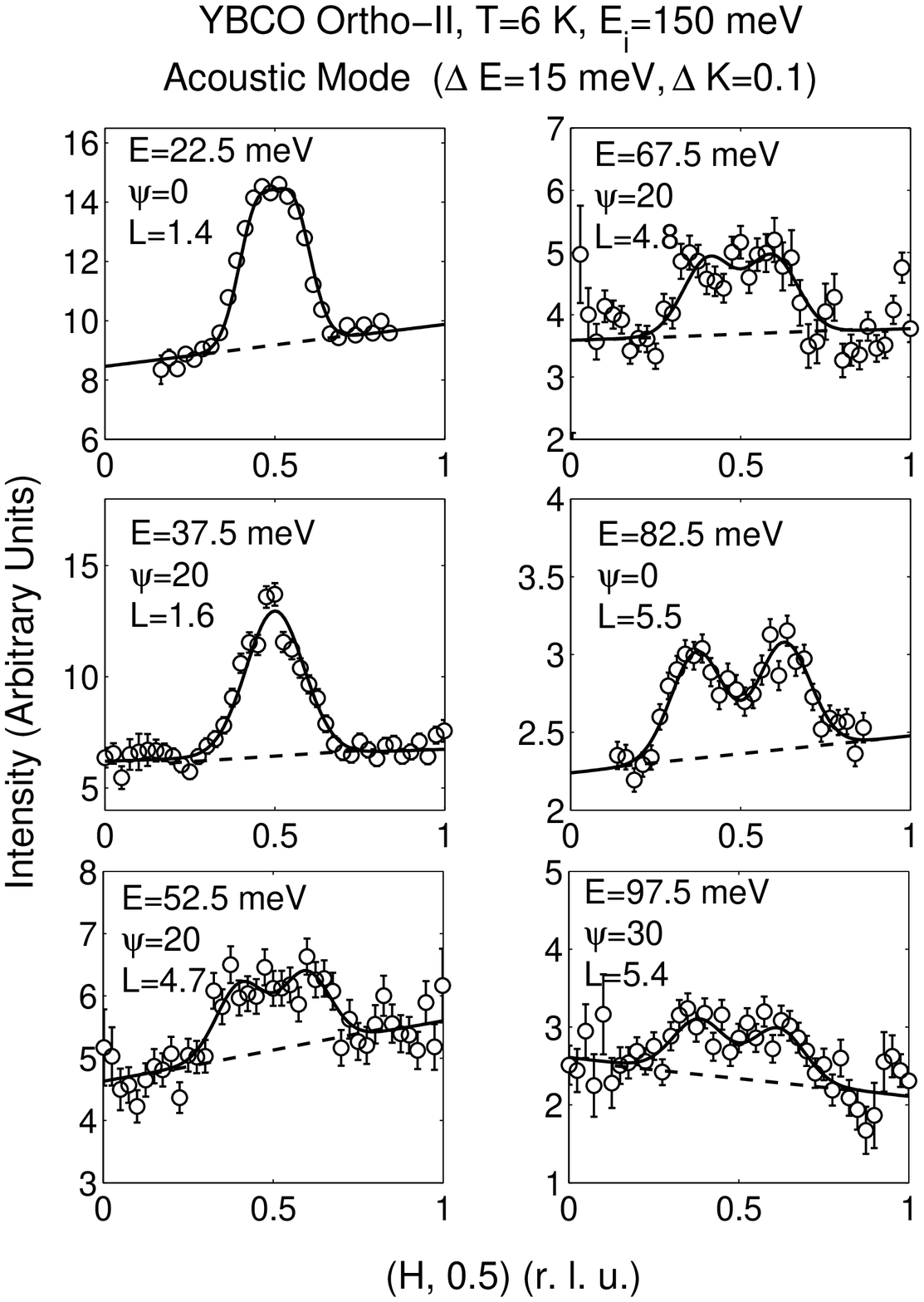}
\caption{\label{acoustic_disp} Constant energy cuts through the correlated peak at the ($\pi$,$\pi$) positions are plotted.  An energy integration of $\pm$7.5 meV was used and the data are integrated $\pm$0.05 r.l.u. along the [010] direction.  The solid lines are guides to the eye. By conducting similar constant energy cuts to those displayed here the dispersion curve was obtained up to $\sim$ 100 meV energy transfer.}
\end{figure}

\begin{figure}[t]
\includegraphics[width=8cm] {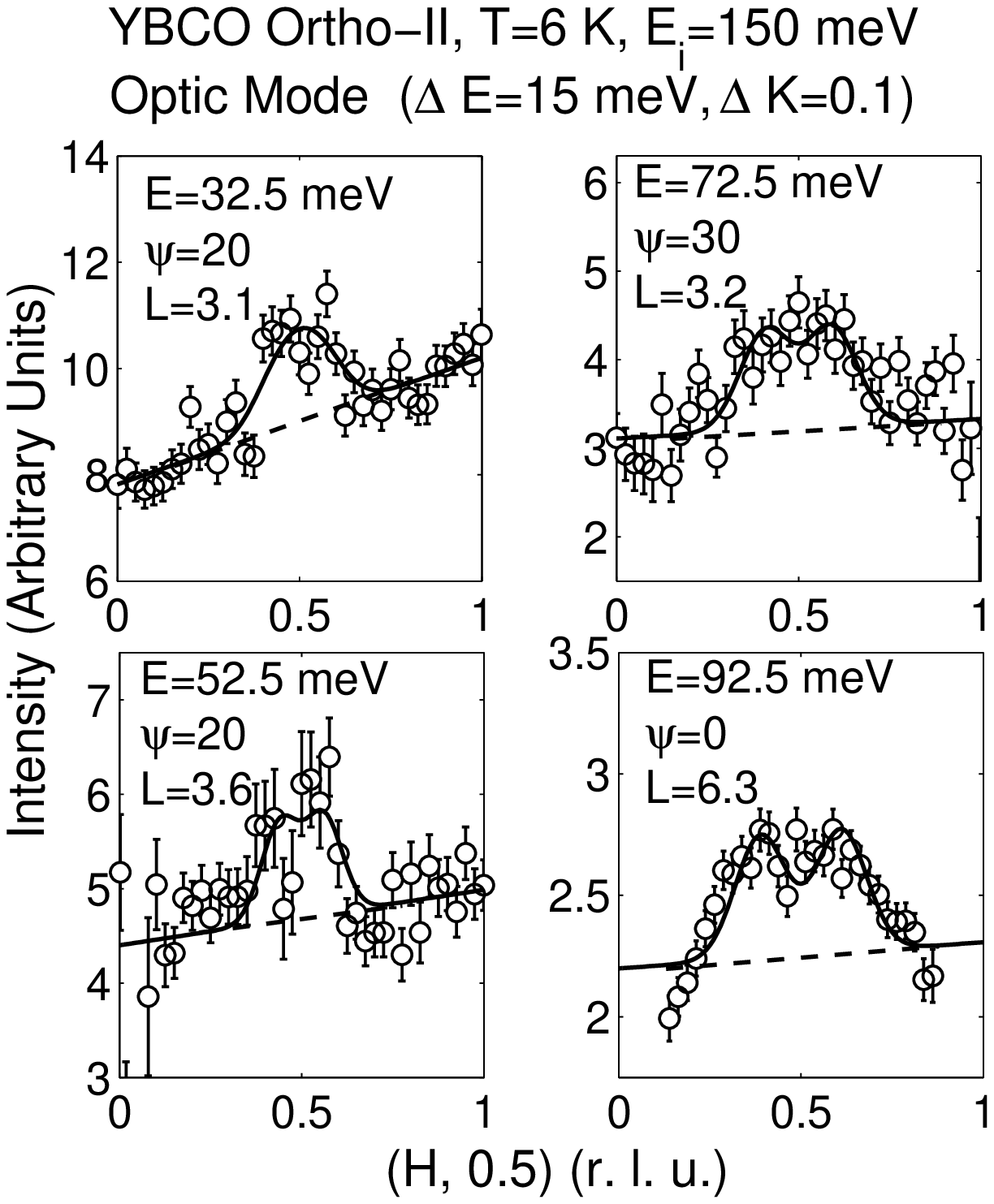}
\caption{\label{optic_disp} Constant energy cuts through the correlated peak at the ($\pi$,$\pi$) positions are plotted.  An energy integration of $\pm$7.5 meV was used and the data are integrated $\pm$0.05 r.l.u. along the [010] direction.  The solid lines are guides to the eye. By conducting similar constant energy cuts to those displayed here the dispersion curve was obtained up to $\sim$ 100 meV energy transfer.}
\end{figure}

\begin{figure}[t]
\includegraphics[width=8cm] {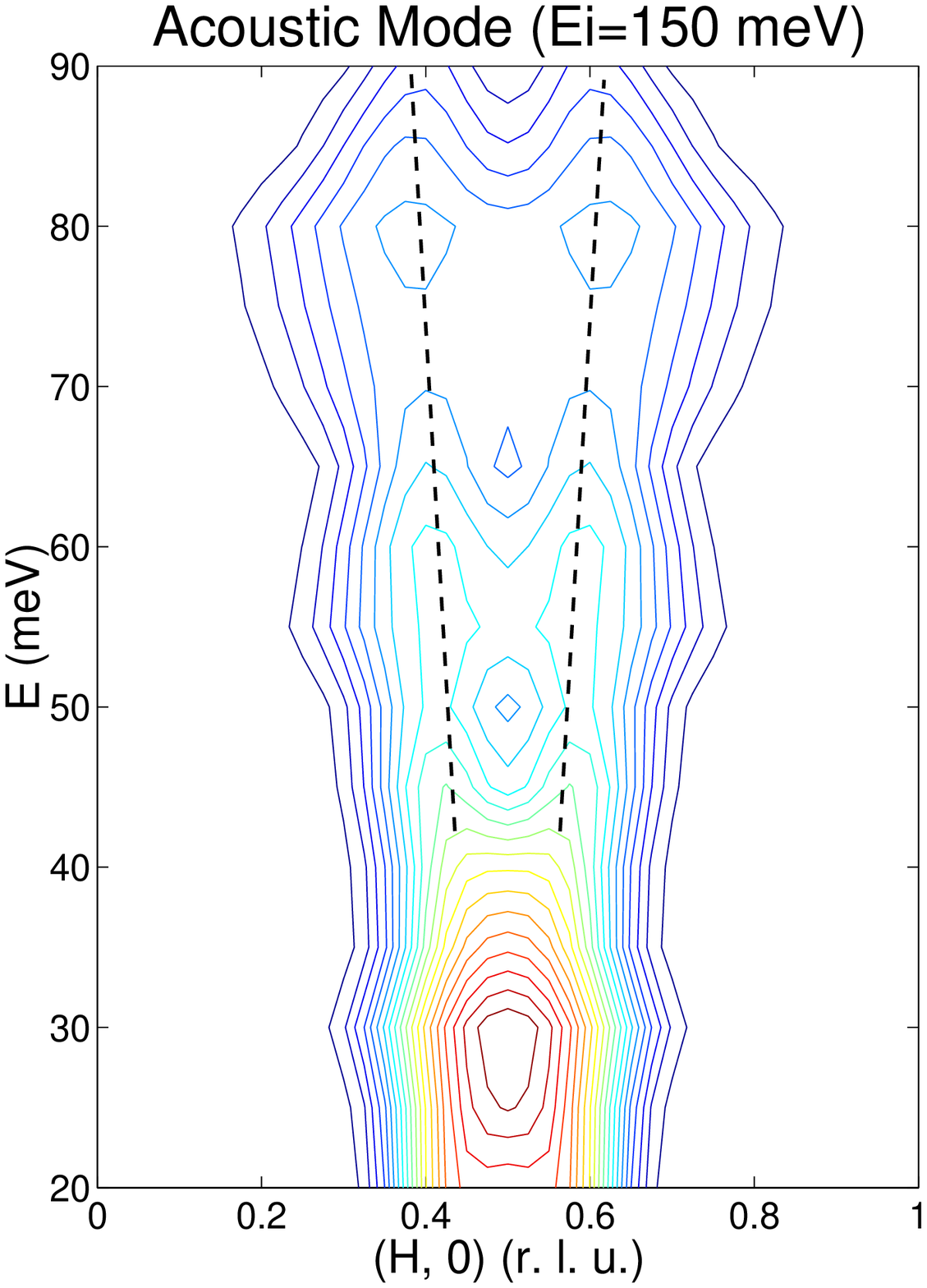}
\caption{\label{constQ} A map of the acoustic intensity (color online) as extracted from Gaussian fits to constant energy scans.  The intensity has been plotted on a logarithmic scale and is in arbitrary units normalized to unity.  From this plot, it can be seen that an energy scan at constant Q, at energies exceeding the resonance energy, will not display a peak or evidence of propagating modes. There is nonetheless evidence of two ridges that move to larger momentum as energy increases and that give a qualitative indication of a characteristic velocity (dotted line).}  
\end{figure}

    To investigate how the lineshape evolves as a function of energy transfer, one-dimensional cuts along the [100] axes at each energy transfer were fitted to two Gaussians displaced equally from the ($\pi$, $\pi$) position to obtain the dispersion for the acoustic and optic modes above the resonance energy.  As a consistency check, we have also performed the same analysis above the resonance but with cuts along [010].  Identical results were obtained.  The results for the overall momentum dependence of the spin fluctuations is illustrated in Fig. \ref{dispersion}.  Unlike the analysis described in the next section, the one-dimensional cuts to investigate the dispersion involve integrating over only a small range perpendicular to the cut direction.  This procedure is schematically illustrated in Fig. \ref{acoustSFandcutintegration} by the integration window represented by Box \textit{B}.  Examples of such one-dimensional cuts are shown in Fig. \ref{acoustic_disp} for the acoustic mode and Fig. \ref{optic_disp} for the optic mode.

    For energies below the resonance we have confirmed that the stronger incommensurate scattering, along [100], previously obtained with a triple-axis spectrometer is consistent with the MAPS data.  The low-temperature results previously obtained using a triple-axis spectrometer (and described elsewhere) are plotted as open circles in the upper panel of Fig. \ref{dispersion}.  The result at T=6 K is plotted in Fig. \ref{dispersion}.  Here we have imposed a cut-off for each mode such that acoustic is defined as any scan with $\sin^{2}(Q_{z}d/2) > 0.8$ and likewise for the optic mode $\cos^{2}(Q_{z}d/2) > 0.8$.  Based on this strict criterion we expect the measured dispersions to reflect the acoustic and optic dispersion accurately.  The data presented in Fig. \ref{dispersion} were obtained for an energy integration width of $\Delta$E=10 meV and an integration along [010] of $\pm$ 0.05 r.l.u.  We have also verified that the overall results remain unchanged for larger energy bins.  The horizontal errorbars on each point are the errors in the $q$ positions, for two branches except at the resonance energy where only one peak is resolved.  The open circles in the upper panel of Fig. \ref{dispersion} are locations of the incommensurate peaks obtained from our previous reactor experiment on the acoustic mode.  The reactor results are discussed in detail in Ref. \onlinecite{Stock04:69}, and hence will not be presented in detail here.  Despite the fact that the energy resolution used in the current experiment is much broader than that used in our reactor work the two results agree very well.  

        The open square symbols in Fig. \ref{dispersion} represent a polar average around the ($\pi$, $\pi$) position obtained for scans done with $\psi$=0, i.e.  with the incident beam parallel to the [001] direction.  For all energy transfers in the range $\hbar \omega$=45 meV to 105 meV, in steps of 5 meV, the polar average was evaluated with respect to the ($\pi$, $\pi$) position.  Points were binned in 0.017 rlu steps from ($\pi$, $\pi$).  For the points in each bin the mean distance from ($\pi$, $\pi$) and the mean intensity were evaluated.  A Gaussian fit to the resulting polar average was used to find the position of the maximum intensity for that energy transfer.  The polar average agrees well with the one-dimensional cuts along [100] and [010] as we expect for an isotropic velocity for the high-energy excitations.

    The momentum dependence of the spin fluctuations as a function of energy transfer qualitatively differs from the scattering measured in early work on La$_{1.86}$Sr$_{0.14}$CuO$_{4}$ which appears to show no appreciable broadening in momentum up to energies of $\sim$ 150 meV.~\cite{Hayden96:76}  Our results seem to suggest that the resonance may be interpreted as the top energy scale of the stripe-like incommensurate features at low-energies.  This is confirmed by a recent experiment on La$_{1.84}$Sr$_{0.16}$CuO$_{4}$ by Christensen \textit{et al.}~\cite{Christensen04:0403439}  They found that the low-energy incommensurate peaks do meet at the ($\pi$, $\pi$) position at $\sim$ 40 meV, a similar energy scale to that found in YBCO.  As suggested in Ref. \onlinecite{Christensen04:0403439}, the similarity between this and the 41 meV resonance measured in optimally doped YBCO may suggest that the resonance is not directly associated with T$_{c}$.  The strong similarity between the low-energy response in LSCO and YBCO may imply that the spin spectrum is common to all cuprates.  The low-energy responses measured in ortho-II YBCO$_{6.5}$ and La$_{1.84}$Sr$_{0.16}$CuO$_{4}$ are also qualitatively similar to the spin spectrum found in the nickelates, a well characterized system known to exhibit charge and spin stripes.~\cite{Bourges03:90,Boothroyd03:67}

    The dispersion measured in YBCO$_{6.5}$ contrasts with that of the insulating compound.  The solid lines in Fig. \ref{dispersion} represent the dispersion measured in insulating YBCO$_{6.15}$~\cite{Hayden96:54}.  For the acoustic modes the lower dashed lines represent a fit to linear spin-waves ($ \hbar \omega \propto c q$) to only the highest energy excitations ($\hbar \omega$ $>$ 40 meV).  For the optic modes, the dashed line is the result of a fit to linear spin-wave theory for bilayers.  We obtain a slope of $\hbar c \sim$ 400 meV $\cdot$ \AA\ for the high-energy acoustic scattering in our well-ordered ortho-II crystal.  The spin-wave velocity observed in superconducting YBCO$_{6.5}$ is considerably reduced from the insulator value of $\hbar c$ $\sim$ 650 meV $\cdot$ \AA.~\cite{Hayden96:54}  It is also considerably less than the value of $\hbar c$ $\sim$ 600 meV $\cdot$ \AA\ measured recently for LBCO.~\cite{Tranquada04:429}  

    The acoustic dispersion around the resonance energy is very similar to that to the \textit{x}-like dispersion observed in La$_{1.875}$Ba$_{0.125}$CuO$_{4}$,~\cite{Tranquada04:429} LSCO,~\cite{Christensen04:0403439} and optimally doped YBCO.~\cite{Reznik03:7591}  This suggests that the \textit{x}-like dispersion is common to all cuprates.  Despite the fact that the dispersion is similar, our results in detwinned and oxygen ordered YBCO$_{6.5}$ show the low-energy excitations to be one-dimensional with incommensurate peaks displaced only along the [100] directions and the high-energy excitations to be essentially symmetric around the ($\pi$, $\pi$) position.  We note that a similar value to that obtained here for the spin-wave velocity was found by Bourges \textit{et al.}~\cite{Bourges97:56} for disordered superconducting YBCO$_{6.5}$ (T$_{c}$= 52 K).  Therefore, the high energy excitations are not much influenced by structural disorder, whereas disorder broadens the intrinsic sharpness of the resonance considerably.~\cite{Stock04:69}  This emphasizes our previous assertion that the high energy excitations may not be that sensitive to the effects of twinning and the orthorhombic anisotropy.

    We note that an important difference between the work presented here and that of previous studies (such as Ref. \onlinecite{Bourges97:56}) is that by combining reactor and spallation data we have been able to resolve the dispersion around the resonance and contrast the high-energy spin excitations with the incommensurate rods at low-energies.  Previously, it was not clear if the dispersion at high-energies extrapolated smoothly (as interpreted in Fig. 3 in Ref. \onlinecite{Bourges97:56}) to $\hbar \omega$=0 or not.  Our results clearly show that the dispersion at high-energies is distinctly separate from the low-energy incommensurate fluctuations.  The separation between the two regions is defined by the resonance energy.  The physical mechanism separating these energy regions is not currently clear, however, the high-energy excitations are of a different character than those observed at low-energies below the resonance.  

    To estimate the in-plane (J$_{\|}$) and out-of-plane (J$_{\bot}$) couplings, we have fitted the acoustic dispersion from the resonance to high energies (ignoring the low-energy incommensurate scattering) to

\begin{eqnarray}
\label{gap_acoustic}
{E_{acoustic}(q)=(\Delta_{ac}^2+\epsilon(q)^2)^{1/2},}
\end{eqnarray}

\noindent where $\epsilon(q)=\hbar c q$ (at low-energies) is the acoustic dispersion predicted by linear spin-wave theory~\cite{Hayden96:54,Tranquada89:40} and an arbitrary gap $\Delta_{ac}$ $\sim$ 33 meV has been set close to the resonance energy.  For the optic mode, we used the gapped form predicted by linear spin-wave theory and found $\Delta_{op}$ $\sim$ 50 meV from the highest -energy optic modes (lower panel of Fig. \ref{dispersion}).  Based on this analysis we have obtained J$_{\|}$$\sim$70 meV and J$_{\bot}$$\sim$ 10 meV. The derived velocity $\sim$ 365 meV $\cdot$ \AA, using the gapped form in Eq. \ref{gap_acoustic}, is close to that obtained by extrapolating the high-energy spin-waves to $q$=0.  

    The use of the gapped form for the acoustic dispersion is physically justified by the fact that the low-energy incommensurate branches meet at the resonance energy therefore suggestive that the low-energy incommensurate scattering and the high-energy dynamics can be treated as different branches.  Also the gapped dispersion results naturally if we consider the ground state to be approximated by dynamic stripes or ladders.~\cite{Barnes93:47}

    It is well known that for the $S$=1/2 square lattice, renormalization factors are important.  To estimate the true exchange constants the renormalization factor Z$_{c}$ has previously been used.  This has been estimated from a $1/S$ expansion by Igarashi to be $\sim$ 1.2.~\cite{Igarashi92:46}  Even though it is only a 20 \% effect, the effect of the renormalization factor means that our measured value of 70 meV translates into $\sim$ 60 meV.  The value of J$_{\|}$ of 60 meV is considerably less than the value of 125 meV found in the insulating compounds.  We conclude that the effect of doping has been to decrease the spin exchange, J$_{\|}$, and concomitantly, the velocity by 50 \%.

     For energies below $\sim$ 40 meV we were not able to separate reliably the two peaks in $q$ associated with the optic mode.   The dashed bars in Fig. \ref{dispersion} represent the full width at half maximum $\Delta q$$_{optic}$ of the optic spin response at $\sim$ 30 meV.  We believe the incomplete spin correlations (see next section) may be responsible for the optic tail below $\sim$ 40 meV, and so the data at low energies should not be fitted to our model dispersion.  If the optic tail is the result of dynamic stripes then the strong similarity between the high-energy dynamics from both the dispersion and the integrated intensity (see next section) suggests that the stripes are weakly pinned and therefore the high-energy scattering can be treated differently than the scattering around, or below, the resonance energy.  Our suggestion of the presence of weakly pinned stripes is consistent with recent resonant x-ray scattering data illustrating a weak hole modulation in the CuO$_{2}$ planes, possibly associated with chain ordering.~\cite{Feng04:xx}  If the stripes are weakly pinned, then the spin fluctuations at low-energies would be expected to have very different properties from those at high-energy.  This provides some support for our analysis in which we extract effective exchange constants from only the high-energy data.

    A prediction of the bilayer stripe model~\cite{Kruger04:1354} is that the optic scattering near the resonance energy will be incommensurate.  It is difficult to discuss the possibility of incommensurability in the optic channel in YBCO$_{6.5}$ because the optic scattering is relatively weak and because we know~\cite{Stock04:69} that the incommensurate wave vector is small, $\delta$ $\sim$ 0.06 r.l.u..  As shown in Fig. \ref{dispersion} the width of the optic scattering below $\sim$ 40 meV does not change measurably when analyzed as a single peak.  However the scan at 32.5 meV in Fig. \ref{optic_disp} a gives a slight hint that the low-energy optic scattering may also be incommensurate with a  $\delta$ similar to that of the acoustic channel.   Further studies at higher doping where $\delta$ is larger may be useful in resolving this issue.   

    Although constant energy cuts show well defined peaks, we find that constant-Q cuts up to $\sim$ 100 meV do not show any clear sign of propagating modes.  Fig. \ref{constQ} plots the fitted intensity and widths as a function of energy transfer based on constant energy cuts through the data from 20 to 90 meV.  Based on this figure, it can be seen that constant-Q scans do not show evidence for propogating modes (ie. a peak in a constant-Q scan) above the resonance energy.  This may suggest that the spin fluctuations are heavily damped even at high energies but to confirm this our results need to be extended to much higher energies due to the large velocity of the spin excitations in the energy range studied here.  It is interesting to note that despite the fact we observe two well defined peaks in the acoustic channel (Fig. \ref{acoustic_disp}) it is difficult to resolve the two optic modes even at $\sim$ 93 meV (Fig \ref{optic_disp}).  This may indicate that the optic channel is more heavily damped both in momentum and energy, than the acoustic channel.    

\subsection{Extraction of Acoustic and Optic Intensities}

\begin{figure}[t]
\includegraphics[width=8cm] {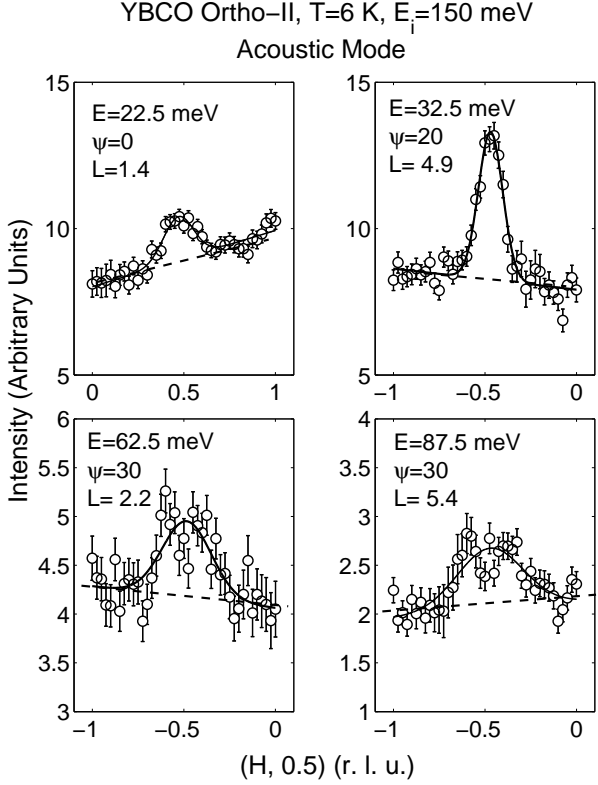}
\caption{\label{acoustic_int} Scans through the acoustic correlated peak for several energy transfers.  The scans have been integrated perpendicular to the scan direction over a broad range as discussed in the text in order to include all of the correlated scattering.  The energy integration width for each scan is $\pm$ 2.5 meV.}
\end{figure}

\begin{figure}[t]
\includegraphics[width=8cm] {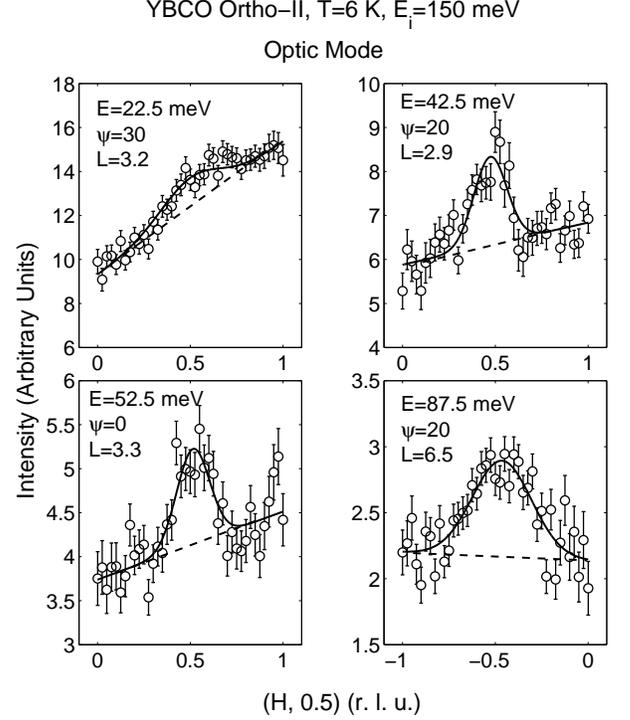}
\caption{\label{optic_int} Scans through the optic correlated peak for several energy transfers.  The scans have been integrated perpendicular to the scan direction over a broad range as discussed in the text.  The energy integration width for each scan is $\pm$ 2.5 meV.}
\end{figure}

\begin{figure}[t]
\includegraphics[width=8cm] {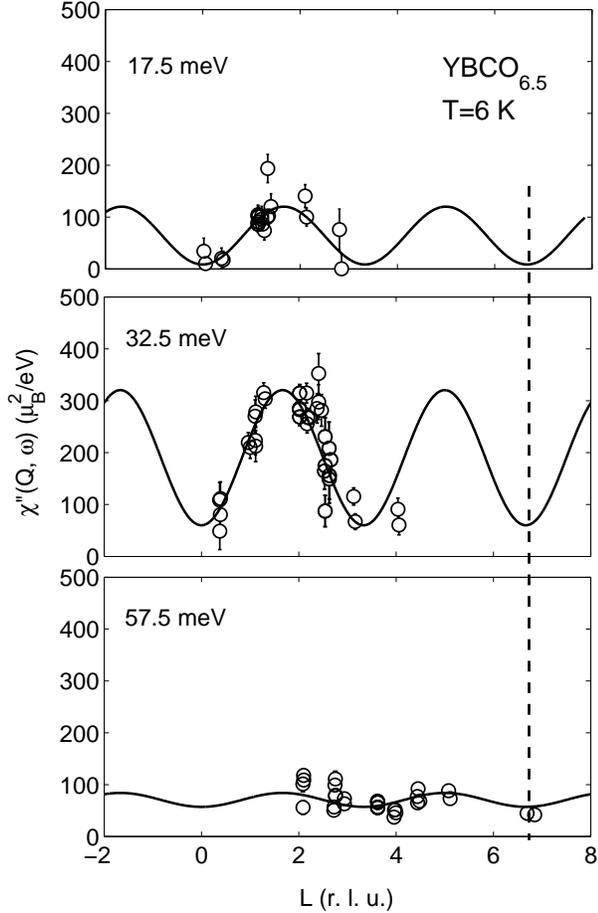}
\caption{\label{L_fit} The intensity as a function of L is plotted at three different energies.  The solid line is the result of a fit to the sum of the acoustic and optic structure factors described in the text.  Based on similar fits, optic and acoustic intensities were extracted as a function of energy transfer as displayed in Fig. \ref{odd_even}.  For the fitted curves, $d/c$ was fixed at 0.3.The dotted vertical line shows where we scanned using a triple-axis to confirm the presence of optic spectral weight at low-energies.}
\end{figure}

\begin{figure}[t]
\includegraphics[width=8cm] {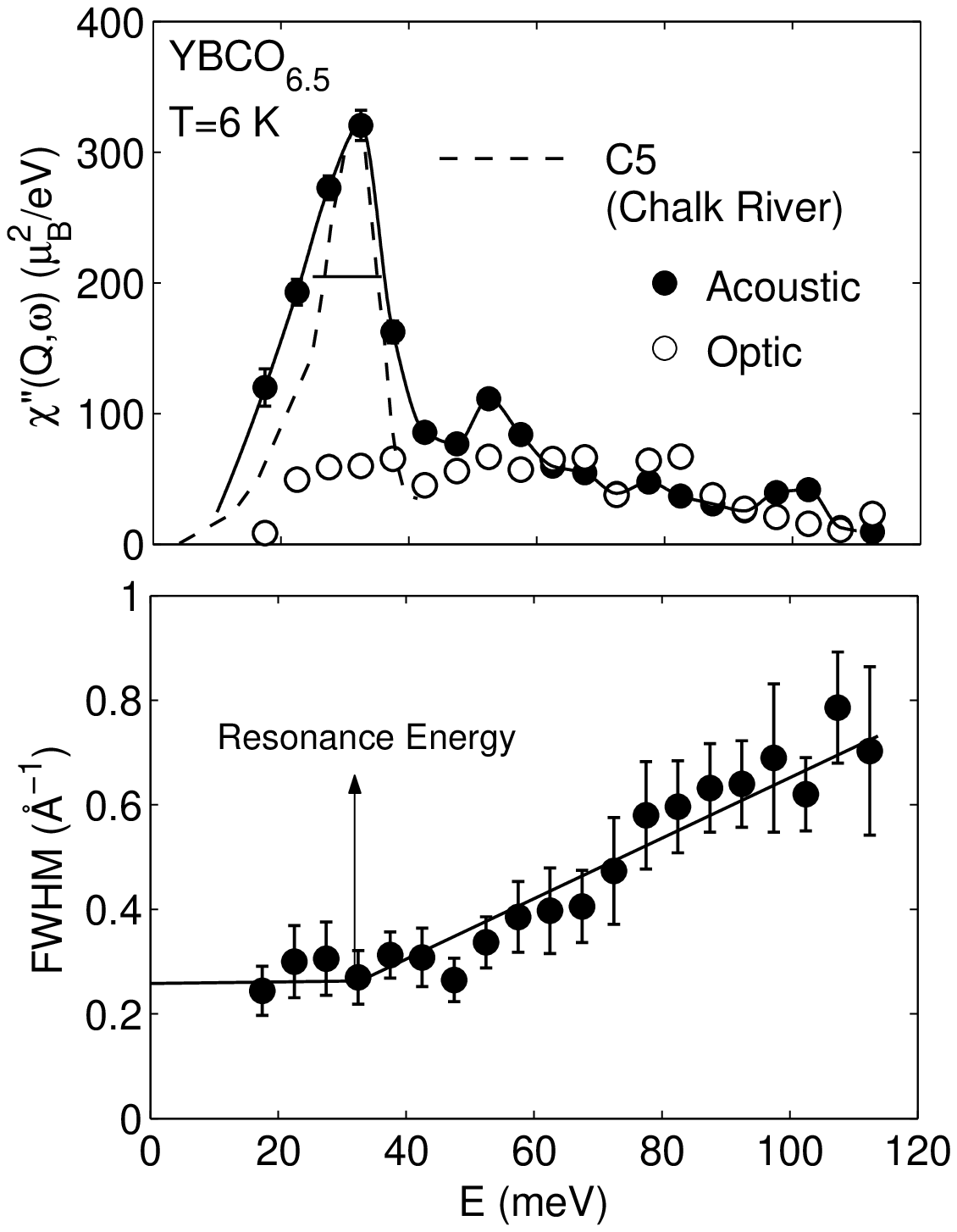}
\caption{\label{odd_even}  The intensities of the acoustic and optic modes are plotted as a function of energy transfer at T=6 K in the upper panel.  The solid bar represents the FWHM resolution of MAPS measured at $\hbar \omega$=0.  The acoustic and optic intensities are roughly equal at high-energy transfers.  The optic mode can be seen to extend to very low-energies compared with the parent antiferromagnetic insulator. The solid lines are guides to the eye.  The dashed line is the low-temperature lineshape of the resonance obtained using a triple-axis spectrometer and scaled down from $\sim$ 1400 $\mu_{B}^{2}/eV$ to 300 $\mu_{B}^{2}/eV$ to account for the poorer energy resolution of the MAPS data.  The lower panel shows the energy dependence of the average FWHM of one-dimensional cuts along H and K for which the transverse momentum integration (along K and H) is set large enough to cover the diameter of the spin response.}
\end{figure}

    The previous sections have illustrated that the lineshape of the magnetic scattering varies dramatically with energy transfer.  When comparing relative spectral weights between the acoustic and optic modes, it is desirable to be able to compare each energy equally without need for detailed lineshape corrections.  In this section we discuss our analysis of the acoustic and optic weights by integrating over one dimension in $q$.

        To analyze the magnetic intensity into its optic and acoustic components as a function of energy transfer we have assumed a symmetric two-dimensional Gaussian line shape (along [1, 0] and [0, 1]) and integrated over a broad range in $\mid$$\bf{q}$$\mid$.  The range is chosen wide enough to collect intensity from all the incommensurate features.  This procedure is illustrated in Fig. \ref{acoustSFandcutintegration} by the integration window represented by Box \textit{A}, which shows a broad integration along the [0, 1] direction perpendicular to the cut direction.  By conducting a one-dimensional cut along [1, 0], with the intensity integrated along [0, 1], we obtained a single peak that includes all of the correlated scattering.  The single correlated peak was fitted to a Gaussian plus a sloping background.  The width of the integral along the [0, 1] was taken to be $\pm$ 2$\sigma$ of the fitted Gaussian along [1, 0].  We repeated the analysis by doing a cut along [0, 1] and integrating by $\pm$ 2$\sigma$ along [1, 0] and obtained consistent results.  Examples of this analysis are displayed in Figs. \ref{acoustic_int} and \ref{optic_int} for scans of predominantly acoustic and optic weight respectively.  The integrated intensity profiles at all energies are reasonably described by a single Gaussian, and more importantly, even when some of the underlying two-peaked structure remains visible, the Gaussian does a good job of estimating the integrated intensity needed as input to the ensuing the acoustic-optic analysis.

    The large integration width perpendicular to the scan direction has smeared out the ring of scattering (as shown in Fig. \ref{colour}) and allows the correlated intensity to be well described by a single Gaussian from which an amplitude can be extracted.  The use of a one-dimensional integral to study the intensity as a function of energy transfer has the advantage of removing any changes in the line shape with energy transfer (as discussed in the next section).  With this analysis we can treat all energy transfers on an equal footing in terms of a single symmetric Gaussian, and study the acoustic and optic components without need for incorporating detailed changes in the line shape.  We discuss the two-dimensional integral as a function of energy transfer later. 

    To extract the acoustic and optic weights at each energy transfer, the amplitude from the Gaussian fit was corrected for the integration perpendicular to the scan direction and the anisotropic magnetic form factor.  We emphasize that this analysis relies on the assumption of a symmetric Gaussian centered at the ($\pi$, $\pi$) position.  The measured peak intensity of a cut along a particular direction through the peak of a two-dimensional Gaussian will depend on the integration window perpendicular to the cut direction.  However, once the width of the Gaussian is known (defined by $\sigma$), the correction for the finite integration perpendicular can be computed to obtain the true peak intensity of the Gaussian.  To check this analysis, several integration widths were chosen and consistent results were obtained.  By plotting the corrected intensity as a function of the different L values we obtain the modulation data shown in Fig. \ref{L_fit}.  The depth of the modulation reflects the acoustic-optic ratio.  The known structure factors,

\begin{eqnarray}
\label{sfactors}
{\chi''(Q_{z}, \omega)} = \chi''_{ac.}\sin^{2}(Q_{z}d/2)+ \chi''_{op.}\cos^{2}(Q_{z}d/2),
\end{eqnarray}

\noindent then enable the acoustic ($\chi''_{ac.}$) and optic ($\chi''_{op.}$) susceptibilities to be obtained from a least-squares fit to the data shown in Fig. \ref{L_fit}. In the fits we have fixed the parameter $d$ to be the known bilayer spacing~\cite{Jorgensen90:41} as successfully done at 33 meV in our earlier reactor experiment.  We emphasize that the separation of spectral weight into acoustic and optic components is solely based on the fact that there are two coupled layers per unit cell, and does not depend on correlations or damping of the two modes.

\begin{figure}[t]
\includegraphics[width=8cm] {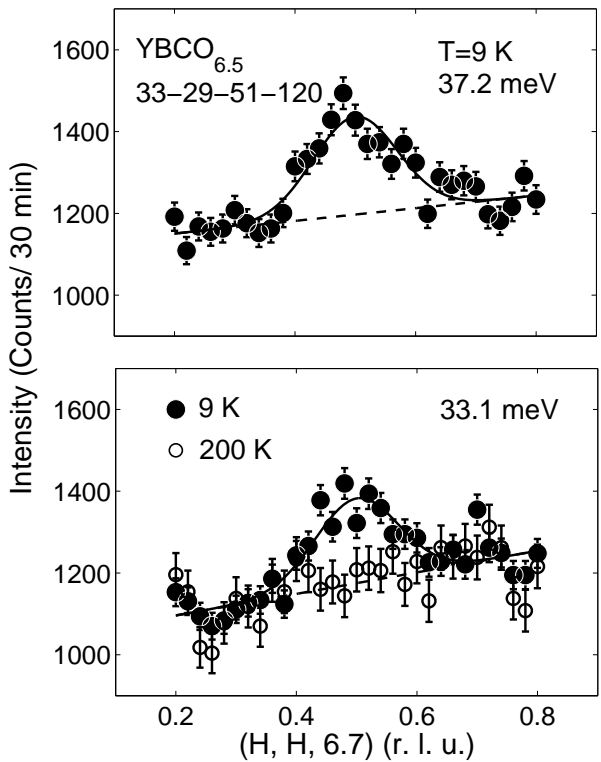}
\caption{\label{nru_optic}  Scans of the optic spin response at L=6.7 with the DUALSPEC triple-axis spectrometer at NRU reactor, Chalk River. It can be seen that optic intensity is still present at relatively low energy transfers.  The solid lines are Gaussian fits to the data.  The solid and open circles represent data taken at 9 K and 200 K, respectively.  The 200 K has had 200 counts subtracted to allow a direct comparison to the 9 K data.}
\end{figure}

    Fig. \ref{L_fit} displays the L dependence of the scattering at three energy transfers.  because the L-dependence of the data extends to relatively low L, we have checked that the optic/acoustic ratio is insensitive to the precise value of $d/c$.The solid line is the result of a fit to Eq. \ref{sfactors}.  At low energies the scattering can be seen to be predominately acoustic while at high energies (above $\sim$ 60 meV) the acoustic and optic modes contribute nearly equally at all L values.  The nearly equal contribution for optic and acoustic modes is consistent with the results of Pailhes \textit{et al.} at 54 meV in YBCO$_{6.85}$.~\cite{Pailhes04:3609}  The fact that we clearly see scattering out to L$\sim$ 7 suggests that the scattering is predominately from Cu$^{2+}$ spins.  Some other forms of magnetism, like the orbital currents described in Ref. \onlinecite{Lee03:68}, would only contribute significantly at small values of L due to their steep form factor decrease.  

    The results for the absolute intensities of the acoustic and optic modes are summarized in Fig. \ref{odd_even}.  The absolute intensity as a function of energy transfer is qualitatively consistent with data published previously for a similar oxygen concentration.~\cite{Dai99:284}  The strong asymmetry with a sharp cut-off at 35 meV that was observed in our reactor data (broken line in Fig. \ref{odd_even}) is observed less clearly because of the broader energy resolution of MAPS near the resonance.  The lower panel of Fig. \ref{odd_even} shows the FWHM, averaged over direction and zone, of a fit of a single Gaussian to the one-dimensional cuts at constant energy described previously.  This figure is a qualitative picture of the dispersion already presented earlier in this paper.  The FWHM gives the overall Q-width of the incommensurate scattering at low-energy, the width of the commensurate resonance at the 33 meV resonance energy, and at larger energies gives a qualitative measure of how the cone of spin excitations opens with increasing energy (discussed in detail in the previous section).  The integrated data presented here (peak and FWHM) are used in the next section to compute the total integral overall momentum and energy. 

    The acoustic fluctuations exhibit the strong resonance at 33 meV seen in our reactor experiments and also exhibits a weaker second peak at around 55 meV. This weak second peak may be consistent with the shoulder previously observed by Bourges \textit{et al.}~\cite{Bourges97:56} above the resonance energy.  We note that a recent phonon study by Pintschovius \textit{et al.}~\cite{Pintschovius04:69} has found intense and nearly dispersionless branches which intersect with the zone boundary in this energy range.  Therefore, we cannot rule out that any extra scattering may be the result of a significant phonon contribution in the range 50-60 meV.  We note that we do not observe any new resonant feature above the 33 meV resonance, apart from the weak 55 meV feature.  Such a resonance has been suggested in theory to result from an antibound state of two holes, the $\pi$ particle.~\cite{Tchernyshyov01:63}  If this new resonance is of a similar strength to that of the 33 meV resonance, our results put a lower bound of $\sim$ 100 meV for the energy of any such resonant feature.  Lee and Nagaosa~\cite{Lee03:68} have predicted that orbital current fluctuations cause additional scattering above the main resonance.  Due to the form factor associated with orbital currents such scattering should only be apparent for small values of L, not accessible in our experimental configuration.  It is currently not clear if orbital currents are contributing to the scattering we observe around 50-60 meV.  

    A particularly interesting result displayed in both Figs. \ref{L_fit}, \ref{odd_even}, and \ref{optic_int} is the presence of optical intensity at relatively low-energies.  Our measurements show that if an optic gap exists, its energy is less than $\sim$ 20-25 meV.  Surprisingly, this is much lower than the optic energy gap of $\sim$ 75 meV measured in the insulating YBa$_{2}$Cu$_{3}$O$_{6.15}$.~\cite{Hayden96:54}

    We have confirmed that the optical response extends to low energies by direct measurement with triple-axis spectrometer where both L and $\hbar \omega$ can be tuned to the desired values.  Scans at the minimum of the acoustic bilayer structure factor, L=6.7, are shown in Fig. \ref{nru_optic} at energy transfers of 33.1 and 37.2 meV and at temperatures of 9 K and 200 K.  The fact that the scattering decreases when we heat to 200 K supports the interpretation that the scattering is magnetic and not from phonons.  The scans confirm our findings from MAPS that a correlated peak of optic symmetry exists at low energies at low temperatures.  These are the same optic fluctuations that extend to 25 meV in Fig. \ref{odd_even}.  A further consistency check is that at 33 meV the ratio of optic to acoustic weight obtained in our reactor experiment is consistent with that found in the MAPS data.  The absolute intensity from the reactor data at 33.1 meV and 37.2 meV was estimated to be 142 $\pm$16 $\mu_{B}^{2}/eV$ and 164 $\pm$ 15 $\mu_{B}^{2}/eV$ respectively based on a phonon calibration.  This is compared with a peak value of $\sim$ 1400 $\mu_{B}^{2}/eV$ previously measured for the acoustic resonance at 33 meV.  The factor of about 10 reduction from the acoustic resonance to the optic weight is similar to that measured at MAPS.  The correlated peak observed in our reactor experiment cannot arise from the feed through of the acoustic resonance at both energies, for we have previously demonstrated that the acoustic weight decreases substantially from 33 to 37 meV.  However, both our MAPS and DUALSPEC data shows the optic weight to be comparable.  Also, previous measurements of the resonance along [001]~\cite{Pailhes04:3609} suggest that at L=6.7, there is essentially no acoustic contribution to the scattering (or at least an attenuation factor much larger than 10).   

    The Chalk River and MAPS data show that the optic fluctuations do not display any resonant behavior around 33 meV unlike the strong acoustic resonance.  A similar non-resonant optic response was obtained by Fong \textit{et al.} in underdoped YBCO$_{6+x}$ who observed that the onset of the optic mode energy ($\sim$30 meV in YBCO$_{6.7}$) declined with doping.~\cite{Fong00:61}  Our results indicate the gradual build-up of non-resonant optic intensity around the resonance energy as has been observed for nearly optimally doped YBCO$_{6.85}$ near 53 meV and in Y$_{0.9}$Ca$_{0.1}$Ba$_{2}$Cu$_{3}$O$_{7}$.~\cite{Pailhes03:8394, Pailhes04:3609}

    The presence of optic weight at low-energies is surprising when viewed in the light of linear spin-wave theory which predicts a large optic gap of order $\sim$ 70 meV for the antiferromagnetically ordered insulator.  It is also a departure from models for the superconductor, such as the renormalized mean field theory of Brinckmann and Lee~\cite{BrinckmannLee00:65}, which predicts an optic gap equal to or larger than the acoustic resonance energy.  In the previous section, we found that the measured high-energy dispersion requires a finite bilayer coupling of order 10 meV, which should have served to preserve the antiparallel spin correlations between layers. The presence of optic spectral weight below the acoustic resonance, however, leads to the idea that there is a reduction in the correlation of the spins between the two layers.  In the limit of strong disorder the correlation length along [001] is less than the bilayer spacing and there would be no L-dependence of the scattering, analogous to the monolayer system.  For weak disorder the correlation length would exceed the bilayer spacing and the scattering would depend strongly on L. For YBCO an intermediate situation prevails. We speculate that the presence of optic spectral weight at low-energies signifies that the spin corelations in each layer are becoming increasingly weakened with increasing doping, hence reducing the modulation in their acoustic/optic spin amplitudes.  Strong decorrelation of the spins along [001] with increasing hole doping has been observed in both the La$_{2-x}$Sr$_{x}$CuO$_{4}$ and La$_{2}$CuO$_{4+y}$ systems in the elastic channel.~\cite{Wakimoto61:00,Lee60:99}  The increasing independence of the planes further highlights the magnetic frustration introduced by doping.  With increasing doping, as the layers become decorrelated magnetically, in terms of the superconductivity the material becomes more strongly coupled.    

    Another explanation for the presence of optic weight at low-energies has been outlined in recent theoretical work based on a bilayer stripe model by Kruger and Scheidl.~\cite{Kruger04:1354}  They considered three different stripe structures based on the position of the rivers of charge in one layer with respect to the other.  For each structure they found that a finite bilayer coupling would introduce optic spectral weight at energies much lower than the optic gap of the parent insulator and quite close to the resonance energy of the doped system (defined here as the energy at which the incommensurate branches of spin excitations meet). Such an interpretation could explain the apparent softening of the optic mode with increasing doping and the presence of a weak resonance in the optic channel.~\cite{Pailhes03:8394} We note that these calculations are based on an ordered magnetic ground state and therefore do not predict the large suppression of spectral weight at low-energies which occurs for non-zero doping, nor the enhancement of spectral weight around the resonance energy.  The prediction of the bilayer stripe model, that the optic scattering would be incommensurate, has been discussed in relation to our data in the previous section.

\section{Integrated Intensity and the Sum Rule}

    One of the surprising results from our reactor study of the low-energy scattering was that remarkably little spectral weight resides at energies less than or equal to the resonance energy.  We found that the total spectral weight made up only about 3\% of that expected from the total moment sum rule.  If the integrated spectral weight were distributed equally over all energies, as it is in the insulating antiferromagnet~\cite{Shamoto93:48}, up to the peak of the dispersion at $\sim$ 250 meV, then we would have expected to have found 40 meV/250 meV $\sim$ 16\% of the total spectral weight.  The presence of so little spectral weight at low-energies necessitates a detailed evaluation of the integrated intensity at energies above the resonance energy.

\begin{figure}[t]
\includegraphics[width=8cm] {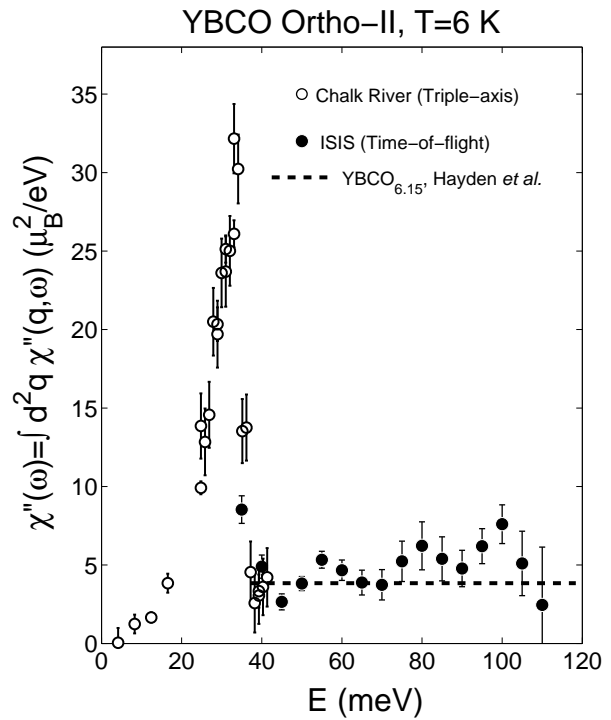}
\caption{\label{integral} Momentum integrated susceptibility as a function of energy transfer for the acoustic mode only.  The integration for the MAPS data (E$_{i}$=150 meV) is plotted above 35 meV with filled circles.  The open circles are data taken previously at Chalk River using a triple-axis spectrometer (E$_{f}$=14.5 meV).  The low-energy spectrum is dominated by the resonance peak at $\sim$ 33 meV.  The high-energy integrated susceptibility is nearly constant and is similar to that previously measured in the YBCO$_{6.15}$ insulator.}
\end{figure}

    To get a qualitative idea for the amount of spectral weight we observe, and the fraction of the total moment which resides around the resonance energy, we consider the sum rule for two \textit{uncoupled} CuO$_{2}$ layers.  Ignoring the effect of the chains, and therefore taking only two Cu$^{2+}$ ions per formula unit, the total spectral weight integrated over all energy and momentum should be given by the total moment sum rule:

\begin{eqnarray}
\label{sum_rule_S} \int d\omega\int d^{3}q \ {S(\bf{q}, \omega)}=2
\times {2 \over 3} S (S+1) g^{2} ,
\end{eqnarray}

\noindent  where $\int d^{3}q$ is the momentum space integral over the correlated peak.  In this equation the factor of 2 comes from the fact that we are taking two Cu$^{2+}$ ions per formula unit.  Here we have assumed that $S(\bf{q}, \omega)$ has been corrected for the bilayer structure factor and the Cu$^{2+}$ anisotropic form factor.  We do expect the total moment sum rule to be reduced as some of the spins will be inevitably destroyed by the doped holes.  Since the doping concentration is so small, \textit{p} $\sim$ 0.1 in the case of Ortho-II YBCO$_{6.5}$, we expect the total moment sum rule to be approximately obeyed.  Equating the above integral to the cross-section for paramagnetic scattering

\begin{eqnarray}
\label{sum_rule_chi}
I \equiv \pi^{-1} \int d\omega \int d^{3}q  [n(\omega)+1]{\chi''(\bf{q}, \omega)}= \\
\nonumber {2 \over 3} \mu_{B}^{2} g^{2} S (S+1).
\end{eqnarray}



\noindent For S=$1 \over 2$ and $g=2$ this gives a total integral of I$=$ $2
\mu_{B}^{2}$.  This estimate provides a useful benchmark with which we can compare our integrated intensities. 

    By integrating the data from 0 to 120 meV in both momentum and energy over the optic and acoustic modes, we find a total integral of $\pi^{-1}\int d^{2}q \int d\omega \ [n(\omega)+1] (\chi''_{ac.}(\bf{q},\omega)+\chi''_{op.}(\bf{q},\omega))$  $\sim$ 0.3 $\mu_{B}^2$ or about 15 \% of the total weight predicted by the sum rule.  The integral from 40 to 120 meV then gives $\sim$  0.23 $\mu_{B}^2$.  As a consistency check we have integrated the acoustic mode from 0 to 40 meV and have found a total spectral weight of $\sim$ 0.07 $\mu_{B}^2$ which agrees reasonably well with the total integral of 0.05 $\mu_{B}^2$ measured in our reactor experiment.  The difference between the reactor and spallation absolute unit calibration illustrates that the experimental error in the measurements quoted here is about 20\% or $\sim$ 0.015 $\mu_{B}^2$.  

    To connect our data with other experiments and theory, we must correct the absolute measurement for the quantum renormalization factor $Z_{\chi}$. $Z_{\chi}$ has been measured from the momentum-integrated spectral weight in both insulating YBCO and LSCO and agrees with the calculated value of $\sim$ 0.5.~\cite{Hayden96:54,Bourges97:79, Coldea01:86}  Therefore, the actual fraction of the total moment measured is about 15 \% of the total moment.  In the insulator YBCO$_{6.15}$, performing the same integral up to 120 meV one gets $\sim$ 0.2 $\mu_{B}^{2}/eV$.~\cite{Hayden96:54}  It is not surprising that we obtain a small fraction of the total moment at these energies as $\chi$$'$$'$ increases dramatically as the dispersion reaches its maximum near the antiferromagnetic zone boundary.

      A simple prediction of linear spin-wave theory for a square plaquette of Cu$^{2+}$ spins is that the low-temperature momentum-integrated susceptibility, $\chi''(\omega)\equiv \int d^{3}q \ \chi''(\bf{q},\omega)$, is independent of $\omega$ in the linear spin-wave region as has been confirmed in insulating YBCO$_{6.15}$~\cite{Hayden96:54, Shamoto93:48}.  As pointed out by Shamoto \textit{et al.}, the momentum-integrated weight is inversely proportional to the exchange constant.~\cite{Shamoto93:48}  Therefore, a reduction in the exchange constant (or spin-wave velocity) will translate into an increase in the momentum integrated spectral weight.  For the insulator, in the low-temperature ordered phase, $\chi''(\omega)$ saturates at a value of $\sim$ 4 $\mu_{B}^2/eV$.  The momentum integral $\int d^{2}q \chi''({\bf{q}},\omega)$ as a function of energy transfer for YBCO$_{6.5}$ is plotted in Fig. \ref{integral}.  The MAPS data are shown for the range from 35-120 meV and our earlier Chalk River reactor data are plotted from 0-45 meV.  The reactor rather than MAPS data are shown below 45 meV because the poorer energy resolution of MAPS for E$_{i}$=150 meV causes the resonance to appear much broader in energy and lower in peak intensity than that measured in our reactor data.  We emphasize, however, that the total integrals for the reactor and time-of-flight measurements are the same within error for energy transfers 0-45 meV.  The two data sets do agree in the energy range of 35-45 meV where the magnetic scattering is smoother in energy.  As illustrated in Fig. \ref{integral}, the prediction for the spectral weight from linear spin-wave theory is violated at low-energies in YBCO$_{6.5}$ where the spin fluctuations are dominated by the intense resonance peak.  

    Because the spectral weight at low-energies has been fully discussed in Ref. \onlinecite{Stock04:69}, we will focus on the spectral weight above the resonance energy.   Fig. \ref{integral} shows that the momentum integral above 40 meV is nearly constant as expected for spin excitations with a linear dispersion.  Its magnitude is consistent with measurements made by Fong \textit{et al}~\cite{Fong00:61} (see Fig. 5).  However, our data provide a strong indication that $\chi$$'$$'$($\omega$) is effectively \textit{constant} well above the resonance energy, unlike previous interpretations of a broad second peak located at a high energy of ~60 meV.  The high-energy acoustic momentum integral is only slightly larger than that (4 $\mu_{B}^2/eV$) measured in the insulator YBCO$_{6.15}$ by Hayden \textit{et al.} at 100 meV.~\cite{Hayden96:54}  This is surprising given that the effective exchange constant, and hence velocity, in YBCO$_{6.5}$ has decreased by a large factor of 60 \%.  This would imply (given the same spectral weight distribution) that the data should approach a constant value of $4/0.6$ $\sim$ 7-8 $\mu_{B}^2/eV$.  Our data in the range 50-100 meV lie in the range 5-7 $\mu_{B}^2/eV$, somewhat smaller than expected but within error.  A small reduction in the renormalization factor Z$_{\chi}$ may possibly occur in the doped system.

    For the total moment, measurements on YBCO$_{6.15}$ have given about $\sim$ 0.2 $\mu_{B}^2$ when integrating up to 120 meV, whereas we find a slightly larger value of 0.3 $\mu_{B}^{2}$.  Because our measured exchange constant is reduced to 60 \% of that of the insulator we expect our measured total spectral weight to be $\sim$ 0.3-0.4  $\mu_{B}^{2}$.   Our result for the total spectral weight is in good agreement with expectations given that the accuracy of the intensity calibration is about 20 \% (as well as the accuracy of the estimated effective exchange constant).  We emphasize that the total integral would be inconsistent, and too small, were it not for the fact that the optic mode softens considerably from the insulator concentration of $x$=0.15 to $x$=0.5.  Therefore, the apparent loss in spectral weight displayed in the acoustic momentum integral is compensated by the presence of optic spectral weight at low energies.  For La$_{1.875}$Ba$_{0.125}$CuO$_{4}$, no correction for any change in exchange constant is required since the spin-wave velocity is similar to that of the insulator.~\cite{Tranquada04:429}  The high-energy momentum integrated spectral weight in La$_{1.875}$Ba$_{0.125}$CuO$_{4}$ is similar to that measured in the insulating monolayer compounds.  In conclusion, we find that the spectral weight at high-energies in YBCO$_{6.5}$ to agree very well with that of the insulator when the change in exchange constant and the softening of the optic mode is taken into account.  

    Despite the fact that the resonance carries a large fraction of the spectral weight at low-energies, it is not significant compared with the total integral extending up to higher energies. This lends support to the idea that the resonance itself cannot be associated with a pairing boson which would require a much larger fraction of the total spectral weight to be centered at the resonance energy.~\cite{Kee02:88} A low spectral weight around the resonance is predicted by the spin-fermion model based on phase space arguments.~\cite{Abanov02:89}
  
\begin{figure}[t]
\includegraphics[width=8cm] {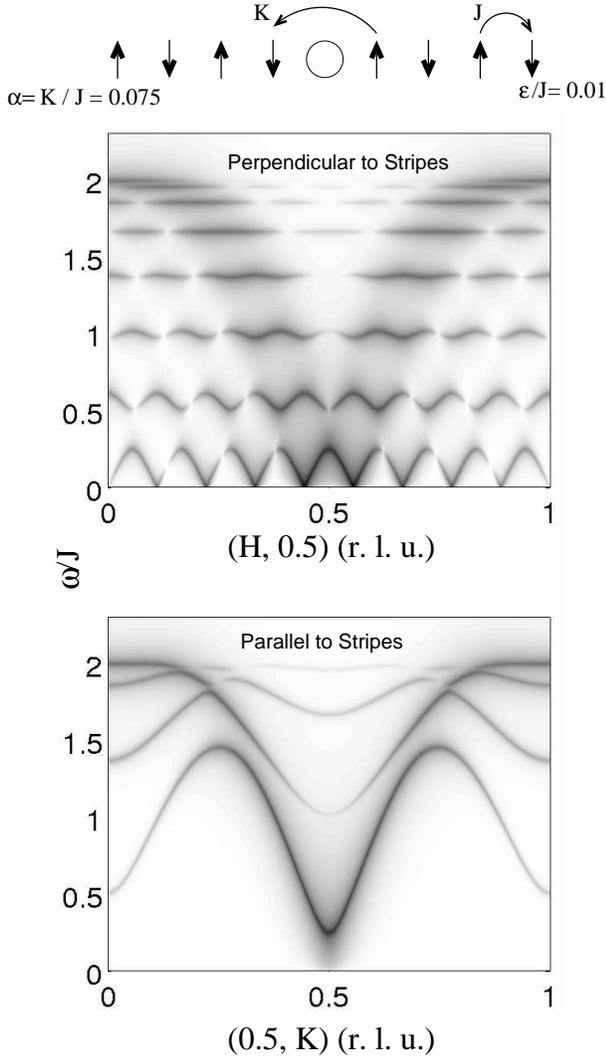}
\caption{\label{gamma_small} (Color online) The spectral weight calculated for an \textit{ordered} spin structure with domains of eight spins separated by a stripe of  $S=0$ states.  The dispersions along the $a^{*}$ direction is shown in the upper and panel and along the $b^{*}$ in the lower panel.  The intensity is plotted on a logarithmic scale.  The value of $\alpha$ and the domain size was chosen to agree with our reactor data, as discussed in the text.}
\end{figure}

    As an intuitive guide we have carried out linear spin-wave calculations based on both an ordered and disordered static stripe ground state (see Appendix).  The hole-rich region is modelled as a vacant site across which a weak exchange coupling acts to produce antiphase domains.  We expect the model to be most appropriate for high frequencies where the excitations see the slow stripe dynamics as essentially static. The stripe spin waves may provide a partial explanation for the presence of reduced spectral weight around the ($\pi$, $\pi$) position.  The stripe model with small damping (Fig. \ref{gamma_small}) shows that significant spectral weight is located throughout the Brillouin zone, in contrast to the conventional spin wave model whose weight is concentrated around ($\pi$, $\pi$).  For large damping, which we speculate may occur for YBCO$_{6.5}$, the extra spectral weight would be distributed throughout the entire zone, would be unobservable as a well-defined correlated peak and so would not contribute to our integral for the spectral weight.  Although this would entail a reduction of weight near ($\pi$, $\pi$), our calculations show that the weight located far from the ($\pi$, $\pi$) position (and below $\hbar \omega \sim$ 120 meV) comprises only $\sim$ 10 \% of the total spectral weight.  Therefore, in terms of spectral weight located around ($\pi$, $\pi$), the stripe model does not predict a substantial change from that of the insulator.

\begin{figure}[t]
\includegraphics[width=8cm] {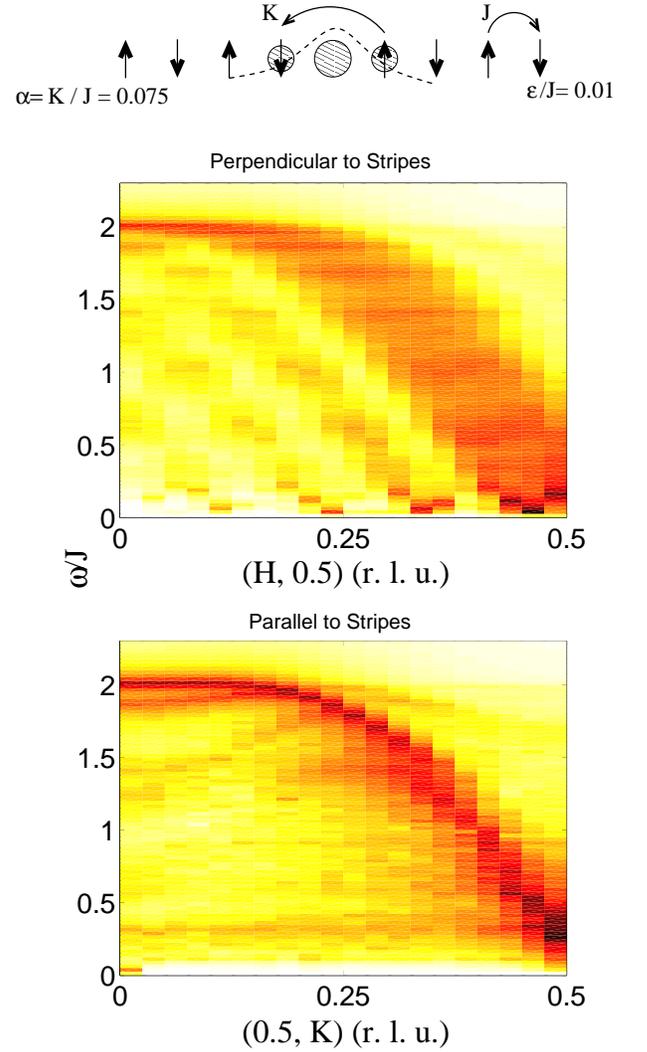}
\caption{\label{disorder} (Color online) The spectral weight calculated for a disordered stripe state on a logarithmic scale.  Instead of the hole always occupying the same position, the hole was allowed to shift randomly along [100] perpendicular to the stripe boundary.  The detailed bands found in the ordered stripe state are found to be smeared out considerably.}
\end{figure}

    The dispersion and distribution of spectral weight at high energies predicted by the stripe calculation shown here (and indeed those presented elsewhere) differ dramatically from the data.  Linear spin-wave theory based on an ordered stripe ground state predicts bands in the high-energy spectrum with substantial gaps of order 0.5 $J$$\sim$ 50 meV below $\hbar \omega$ $\sim$ $J$.  Instead of large gaps in the energy spectrum we observe the spectral weight to be distributed uniformly above the resonance energy.  The predicted spectrum based on an ordered stripe ground-state would also result in an significant anisotropy in the high-energy dynamics.  This is clearly not the case in our measurements as illustrated in Figs. \ref{colour} and \ref{symmetry} and as discussed previously.  Since we know the measured high-energy spin spectra are continuous in energy as well as relatively isotropic (Fig. \ref{dispersion}) we have extended our stripe model to include roughened  stripe boundaries.  This may also simulate stripes moving slowly relative to the high-frequency spin waves.  We have performed calculations where the position of the hole-rich region, represented by a vacancy, is randomly placed to the left of, to the right of, or on the mean position of the boundary in the above 8-spin-plus-vacancy  stripe model.  The results given in Fig. \ref{disorder} show that the disordered stripe model partially eliminates the band-gaps, however there still exists a large anisotropy not observed in our data.  We therefore conclude that only the low-energy dynamics are similar to that expected from an ordered stripe ground state.  

    Although the stripe model correctly predicts incommensurate modulation below the resonance, we emphasize that the model, in particular its spectral weight, is not valid at low energies.  This is because the stripes are dynamic (Ref. \onlinecite{Kivelson03:75}) rather than static in YBa$_{2}$Cu$_{3}$O$_{6.5}$, because the bilayer coupling is assumed small, and because spin destruction by hole doping is ignored.  The model provides only a qualitative way of interpreting the spin map in $q$ and $\omega$ at high energy.

    The spectral weight for a ladder is illustrated in the recent calculations of Uhrig \textit{et al.}~\cite{Uhrig04:2659} who calculated the triplon excitation spectrum associated with a two-leg spin ladder.  Two van Hove singularities were found, around the resonance energy (or when the incommensurate branches meet) and at the top of the spin energy dispersion.  It is likely that a good fraction of the remaining spectral weight exists at higher energies (the high-frequency van Hove singularity) and therefore is not included in our total moment integral.  The calculations of Uhrig \textit{et al.} also show that the momentum-integrated spectral weight is nearly constant between the two van Hove singularities.  This agrees with our observations.  However, their calculations predict the absolute spectral weight at high-energies to be substantially reduced from that in a generic cuprate insulator (like YBCO$_{6.15}$).  This is inconsistent with our results and those measured in other cuprates.  Experiments extending to much higher energies are clearly required to resolve the issue of total spectral weight.

\section{Discussion}

    One of the key observations in this experiment is the presence of excitations at high-energy transfers similar to the spin-waves observed in the parent insulating compounds.  This is based on the high-energy dispersion as well as the absolute momentum integrated spectral weight at high-energies.  The spin excitations are well-defined in wave vector but not in energy.  A direct comparison of these results to those of the low-energy excitations previously studied suggests two distinct regions of the excitation spectrum:  one at low-energies, less than 33 meV, and well described by one-dimensional incommensurate fluctuations, and another at high-energy with excitations characterized by commensurately centered fluctuations that disperse in a similar manner to spin-waves.

    Based on the measured dispersion in La$_{1.875}$Ba$_{0.125}$CuO$_{4}$~\cite{Tranquada04:429} and La$_{1.84}$Sr$_{0.16}$CuO$_{4}$~\cite{Christensen04:0403439}, the low-energy dispersion with incommensurate peaks meeting at higher energies appears to be a common trait of the cuprate superconductors.  However, the momentum integrated acoustic susceptibility as a function of energy transfer we measure in YBCO$_{6.5}$ contrasts with that measured in La$_{1.86}$Sr$_{0.14}$CuO$_{4}$.~\cite{Hayden96:76}  In particular no sharp well-defined resonance peak in the susceptibility is observed in LSCO.  As noted in our previous reactor study, the spectral weight under the resonance comes from the low-energy spectral weight suppressed in the superconducting state.  Therefore, a combination of the a higher energy at which the incommensruate peaks meet and a lower superconducting gap may explain why the resonance peak is not as well defined in LSCO compared with YBCO.
    
    The occurrence of incommensurate scattering that is described by a dynamic stripe model only at low-energies suggests that stripes are weakly pinned in the YBCO$_{6+x}$ system.  This result provides an important energy scale for theories involving low-energy fluctuating stripe phases.~\cite{Zaanen96:53}  The evolution with energy from stripe-like excitations that are hydrodynamic in nature to spin-waves also provides an explanation for the breakdown of $\omega$/T scaling observed at high-energies in our previous reactor work.  We emphasize that even though we use the language of stripes here, any model which results in one-dimensional spin density wave fluctuations at low-energies would be consistent with our data. 

    One way of reconciling these two regions in the excitation spectrum in terms of fluctuating stripes is that the resonance may represent a characteristic energy defined by the size of a stripe domain.  For spin fluctuations with a wavelength greater than the domain size (low-energy excitations) the dynamics of the domain walls will be important.  For high energies where the wavelength is shorter than the distance between domain walls the dynamics of the stripes will not be important and the excitations will look similar to those of the parent insulating compound.  This physical picture is partially reflected in our mean-field calculations (see Appendix) based on a disordered stripe ground state.  This response is qualitatively similar to that of stripe-liquid theory.~\cite{Zaanen97:282}  The experiment shows that the spin response then  shows stripe-like behavior only at low energies.  This then implies, in the context of stripes, that the rivers of charge are only weakly pinned.

    The excitation spectrum of a stripe ground state was first calculated by Batista \textit{et al.}~\cite{Batista01:64} and has recently been extended by Kruger and Scheidl~\cite{Kruger03:67} and Carlson \textit{et al.}~\cite{Carlson04:2231}  These calculations, as well as our own linear spin-wave calculation, show that a stripe ground state cannot exhibit, in the case of weak stripe coupling and disorder (see Appendix), a symmetric cone around the ($\pi$, $\pi$) position.  Despite the fact that disorder helps to remove the gaps, a significant anisotropy still exists in the line width at high-energies when scanned along [10] compared with [01].  This anisotropy results from the intersection of two spin-wave cones emanating from the incommensurate positions.  This is in contrast to the relatively isotropic widths we observe in YBCO$_{6.5}$.  Also, the spin-wave calculation is not capable of explaining the suppression of spectral weight at low-energies, for it ignores the magnetic frustration introduced by doping and as well assumes an ordered ground state.  

    The existence of excitations above the resonance energy, similar to those found in the parent insulating cuprate compounds, suggests that the resonance may be interpreted as a gapped mode which gradually decreases in energy with decreasing doping.  Such an energy gap was predicted in the early theory of Chubukov \textit{et al.}~\cite{Chubukov94:49} for a quantum disordered ground state.  More recent developments by Demler \textit{et al.} suggesting the close promximity of a spin density wave (SDW) phase may explain the existance of the low-energy incommensurate fluctuations.~\cite{Demler01:87}   The magnitude of the energy gap is predicted to decrease according to a power-law as the N$\acute{e}$el state is approached, which is qualitatively seen in the YBCO$_{6+x}$ system. Such a response is also predicted in the spin-fermion model.~\cite{Morr98:81} To test this prediction it would be valuable to study the high-energy excitations near the boundary of antiferromagnetism and superconductivity.

    Predictions for the high-energy dispersion have been made in the random-phase approximation by Brinckmann and Lee~\cite{BrinckmannLee00:65}  and by Kao \textit{et al.}~\cite{Kao00:61}  Brinckmann and Lee predict the scattering to be incommensurate below the resonance and to be commensurate at the resonance energy, consistent with experiment.  The dispersion above the resonance is predicted to have a separation between $+q$ and $-q$ branches in an (H 0 0) scan, that increases with energy, similar to spin waves.~\cite{BrinckmannLee00:65}  In the model by Kao \textit{et al.} the separation in $q$ is constant or slightly declines with increasing energy (see Fig. 1 of Ref. \onlinecite{Kao00:61}), unlike our observation of an opening in the q separation with energy that reflects a characteristic spin velocity.  The high-energy line shape predicted in Ref. \onlinecite{Kao00:61} is nearly isotropic and is consistent with the near symmetric ring observed in this experiment. 

    Despite the several similarities in the high-energy spectrum of the cuprates recently measured and discussed in this paper, there are many key differences which are currently not resolved.  First, despite the fact that we observe a symmetric ring of scattering, recent measurements reported in Ref. \onlinecite{Hayden04:429} show a clear anisotropy around the ($\pi$, $\pi$) position.  Also, despite the fact we observe the momentum integrated spectral weight to be nearly frequency independent above the resonance energy, previous studies (including Ref. \onlinecite{Bourges97:56} and Ref. \onlinecite{Dai99:284}) report a significant peak in the integrated spectral weight at high energies.  Some of these apparent discrepancies may be associated with different dopings as Ref. \onlinecite{Sun04:93} has suggestted the existance of a metal to insulator transition in YBCO$_{6+x}$ at $x$$\sim$ 0.07.  Nevertheless, further studies are required to resolve these discrepancies.

\section{Conclusion}

    We have shown that the spin excitations of an underdoped YBCO superconductor evolve from stripe modulations at low-energies to a near symmetric cone of spin-waves above the resonance energy.  Based on the dispersion we have estimated the in-plane exchange constant to be $\sim$ 75 meV, 40\% less than the exchange of 125 meV in YBCO$_{6.15}$.  In contrast to the insulating compound we find that optic weight is present at relatively low-energies. We suggest that with increasing doping the bilayer spins are gradually decoupled as the superconducting coupling of the carriers increases.  We find that the total spectral weight is similar to that expected from the insulator, when renormalization factors and the reduced exchange constant are taken into account.  The form of the high-energy spectrum cannot be explained with an ordered stripe model, nor one with weak disorder.  The high-energy excitations are similar to spin-waves in the insulator based on both dispersion and spectral weight.  The data are inconsistent with what is expected from a static ladder ground state.  We note that the several similarities of the excitation spectrum we measure in Ortho-II YBCO$_{6.5}$ with that measured in La$_{1.875}$Ba$_{0.125}$CuO$_{4}$ suggests that the spin dynamics in doped CuO$_{2}$ systems may have a universal explanation.

\begin{acknowledgments}

    We thank the late B. Statt, T. Perring, and Z. Tun for useful discussions.  We also thank  L. E. McEwan, M. Potter, A. Cull and R. Sammon for technical support at NRC, Chalk River.  The work at the University of Toronto and the University of British Columbia was supported by the Natural Sciences and Engineering Research Council of Canada and C. Stock acknowledges a  GSSSP supplement from the National Research Council of Canada.

\end{acknowledgments}

\section{Appendix: Excitations of an ordered and disordered stripe ground state}

    The qualitative nature of the spin excitations above and below the resonance energy are very different.  The low-energy excitations are incommensurate with peaks displaced only along the [100] direction, in qualitative agreement with the spectrum predicted for a stripe ground state.  The high energy excitations are \textit{isotropic} in the direction of q, within experimental error, and are qualitatively similar to spin-waves originating from the ($\pi$, $\pi$) position.  Given the different behavior at high and low-energy energies it is important to determine if the two regimes can be understood in terms of the same ground state.  In this section we describe linear spin-wave calculations for an \textit{ordered} stripe ground state as well as numerical simulations describing the effects of disorder on the spectrum.   We find the results of the calculation are not consistent with experiment.

     In our calculations, which parallel those of Batista \textit{et al.}~\cite{Batista01:64}, Kruger and Scheidl~\cite{Kruger03:67,Kruger04:1354}, and Carlson \textit{et al.}~\cite{Carlson04:2231} we have numerically calculated the excitation spectrum of an ordered stripe ground state with one-dimensional domains of 8 spins separated by a vacant spin-site.  The vacant site is taken to model the region of large hole density separating antiphase domains. The size of this domain structure is expected from the known position~\cite{Stock04:69} of the low-energy incommensurate modulation in Ortho-II YBCO$_{6.5}$.  We have used antiferromagnetic Heisenberg exchange between the spins and have varied the strength of the weak coupling between domains across the vacancy at the charge stripe, (\textit{K}), relative to the strong coupling within a spin domain, (\textit{J}), so as to define the key ratio $K/J \equiv \alpha$.  Having fixed the domain size from the position of the low-energy incommensurate peaks the parameter $\alpha$ determines the position at which the low-energy incommensurate rods meet, or where the resonance energy occurs in experiment.  The weak antiferromagnetic exchange, e.g.  $\alpha=0.075$, between domains couples spins two sites apart and so leads to antiphase domains.  We have determined the relative cross-sections as well as the energies so as to compare with experiment.

    The energy spectrum of an ordered spin ground state can be determine from dynamical mean-field theory by solving the set of coupled, first-order equations given by

\begin{eqnarray}
\label{dynamic_mf}
{d_{t}\langle {\bf{S}}(l,t) \rangle={\bf{h}}(l,t) \times \langle {\bf{S}}(l,t) \rangle}.
\end{eqnarray}

\noindent Here, ${\bf{h}}(l,t)$ is the molecular field acting on the spin ${\bf{S}}(l,t)$ caused by its nearest neighbors,

\begin{eqnarray}
\label{molecular field}
{{\bf{h}}(l,t)=\sum_{n_{i}} J(l,l+n_{i}) \langle {\bf{S}}(l+n_{i},t)\rangle}.
\end{eqnarray}

\noindent By assuming a particular ground state structure $\langle {\bf{S}}^{o}(l) \rangle$ and expanding the above equations linearly in deviations about this ground state ($\delta \langle {\bf{S}}(l,t) \rangle$) the following set of linear equations is obtained

\begin{eqnarray}
{-id_{t}S^{+}(l,t)}=\sum_{j}(h^{0}(l)\delta(l,j)-S^{0}(l)J(l,j))S^{+}(j,t)
\end{eqnarray}

\noindent or,

\begin{eqnarray}
{-id_{t}S^{+}(t)}=A S^{+}(t).
\end{eqnarray}

\noindent In this equation $S^{+}(l,t)= \delta \langle S_{x}(l,t) \rangle + i \delta \langle S_{y}(l,t) \rangle $, $h^{0}(l)$ is the molecular field with $\langle {\bf{S}}(l+n_{i},t)\rangle$ set to the ground state values ($\langle{\bf{S}}^{0}(l) \rangle$), and $\delta(l,j)$ is equal to one when $l=j$ and zero otherwise.   By diagonalizing the matrix $A$ the energy spectrum can be found for any ground state and spin-coupling.

    To obtain the spectral weight as a function of momentum and energy transfer we have solved for the Green's function $G(S^{+}(l,t),S^{-}(j,0))=-i\Theta(t)\langle[S^{+}(l,t),S^{-}(j,0)]\rangle$, and then taken the imaginary part to get the neutron scattering cross-section.  Solving the equations of motion for the Green's function,

\begin{eqnarray}
{\sum_{k} (\delta(l,k) \omega + A(l,k)) G(S^{+}(k,\omega),S^{-}(j,0))}= \\
\nonumber \langle[S^{+}(l,0),S^{-}(j,0)]\rangle,
\end{eqnarray}

\noindent the neutron cross-section was obtained by setting $\omega$ to $\omega+i\epsilon$, inverting,  and summing all possible $G(S^{+}(l,\omega), S^{-}(j,0))$.  The parameter $\epsilon$ represents some form of damping.

    The results for the eigenvalues and the spectral weight are shown in Fig. \ref{gamma_small} for a domain size of 8 spins and a ratio of inter- to intra-domain coupling of $\alpha$=0.075.  Taking $J$=125 meV, we have chosen the domain size to give the correct position for the low-energy incommensurate peaks as measured in our reactor data.  The value of $\alpha$ was chosen such that the top of the lowest branch of the dispersion coincided with the position of the measured resonance energy.~\cite{Stock04:69,Batista01:64}  We note that such a detailed microscopic model for the resonance has not been conclusively proven to date.

    The spectrum displays eight branches, as expected given the ground state chosen in this calculation.  The energy axis is given in units of the intra-stripe coupling ($J$).  It is worth noting that the spectral weight is concentrated roughly in the region where excitations would be present in the insulating compound. Thus, we see in Fig. \ref{gamma_small} that the weight lies close to where a single conventional spin-wave band would occur as one moves from one band to another of the multi-branch dispersion of the stripe model.  The concentration of spectral weight around individual branches separated by a finite gap is not observed in the experiment.  It is satisfying, however, that the spin waves of the stripe model do show one-dimensional anisotropy similar to observation: along [0.5, K] a gap occurs at the resonance energy with little weight at lower energy, while along [H, 0], perpendicular to the stripe, the energy falls to zero and the intensity grows on approach to the incommensurate momenta.

    Based on our reactor studies of the low-energy spectrum of Ortho-II YBCO$_{6.5}$, we expect the spin response to be highly damped.  In our previous analysis of the low-energy scattering we estimated a dynamic correlation length $\kappa \sim$ 20 \AA\ and a spin-wave velocity of $\hbar c \sim$ 300 meV $\cdot$ \AA\, giving a large energy broadening of $2\Gamma \sim$ 30 meV.  In terms of the stripe picture this could be taken as evidence for the presence of a large amount of disorder along the stripe of $S$=0 states.  In an attempt to simulate this we have arranged the holes on a finite lattice to have a random and normal distribution (with standard deviation $\sigma$=1 cell or one Cu$^{2+}$ site) around the ordered position such that the hole can be shifted up two unit cells along the [1, 0] direction.  Using periodic boundary conditions we have then calculated the spin spectrum based upon a 1024 spin lattice ground state(32 by 32 spins) using the same formulae presented above.   The results were not found to change significantly for smaller lattices.  The results presented in Fig. \ref{disorder} are averaged over seven different iterations of the 32 $\times$ 32 random-hole structure.

  The results for such a configuration is presented in Fig. \ref{disorder} and shows that the spectral weight can be smeared out considerably by such a disordered ground state.  This calculation also shows that many of the features displayed by the ordered system displaced far from the ($\pi$, $\pi$) position are smeared out.  Also, the spectral weight again appears as a correlated peak around the (Q, $\omega$) positions expected in the absence of stripes and the gapped structure calculated for the ordered system is now smeared out in energy and momentum.  Because the model assumes an ordered ground state it is not able to reproduce the suppression of the low-energy spin fluctuations, which we have previously found to be well described by overdamped spin-waves.  Our formulation of a disordered stripe structure is expected to be more appropriate for higher-energy excitations (above the resonance) as the energy transfer is greater than the broadening of 2 $\Gamma$ $\sim$ 30 meV expected from our reactor data.  We note that the high-energy response found in our disordered stripe model is consistent with the increase with energy of the observed width in q reported for commensurate La$_{1.875}$Ba$_{0.125}$CuO$_{4}$, although our data and calculations show less anisotropy at high energy than the calculations for the spin ladder model.~\cite{Tranquada04:429}

    Even though the low-energy scattering below the resonance may be interpreted in terms of stripes, our numerical calculations show that it is difficult to interpret the high-energy scattering in terms of a stripe ground state.  Our results for an ordered ground state (with no disorder) show large band gaps clearly not observed in the data.  The introduction of disorder does  the band gaps but does not remove a large anisotropy, particularly in the linewidth.   Our  stripe model fails to account for the peak in the spectral weight at around 55 meV.  The model does not account for the asymmetric resonance spectral profile found in our previous reactor work.~\cite{Stock04:69}



\begin{thebibliography}{99}
\bibitem{corr} The correlation length was obtained from a Lorentzian fit to the data ($1/(1+(q/\xi)^2)$).
\end{thebibliography}
\end{document}